\documentclass[a4paper,fleqn,usenatbib]{mnras}

\usepackage[T1]{fontenc}
\usepackage{ae,aecompl}

\usepackage{graphicx}	
\usepackage{amsmath}	
\usepackage{amssymb}	
\usepackage{bm}	
\usepackage{natbib}
\usepackage{txfonts}

\def \etal {et~al.~}

\newcommand{\hMpc}{{\ifmmode{h^{-1}{\rm Mpc}}\else{$h^{-1}$Mpc}\fi}}
\newcommand{\hkpc}{{\ifmmode{h^{-1}{\rm kpc}}\else{$h^{-1}$kpc}\fi}}
\newcommand{\kpc}{{\ifmmode{ {\rm kpc} }\else{{\rm kpc}}\fi}}
\newcommand{\kms}{{\ifmmode{ {\rm km\,s^{-1}} }\else{ ${\rm km\,s^{-1}}$ }\fi}}
\newcommand{\hMsun}{{\ifmmode{h^{-1}{\rm {M_{\odot}}}}\else{$h^{-1}{\rm{M_{\odot}}}$}\fi}}
\newcommand{\Msun}{{\ifmmode{{\rm M}_{\odot}}\else{${\rm M}_{\odot}$}\fi}}
\newcommand{\Mhalo}{{\ifmmode{M_{\rm halo}}\else{$M_{\rm halo}$}\fi}}
\newcommand{\Rvir}{{\ifmmode{R_{\rm vir}}\else{$R_{\rm vir}$}\fi}}
\newcommand{\Mstar}{{\ifmmode{M_{\star}}\else{$M_{\star}$}\fi}}
\newcommand{\Vrot}{{\ifmmode{V_{\rm rot}}\else{$V_{\rm rot}$}\fi}}
\newcommand{\ltsima}{$\; \buildrel < \over \sim \;$}
\newcommand{\gtsima}{$\; \buildrel > \over \sim \;$}
\newcommand{\lsim}{\lower.5ex\hbox{\ltsima}}
\newcommand{\gsim}{\lower.5ex\hbox{\gtsima}}

\def\lesssim{\mathrel{\hbox{\rlap{\hbox{\lower4pt\hbox{$\sim$}}}\hbox{$<$}}}}
\def\gtrsim{\mathrel{\hbox{\rlap{\hbox{\lower4pt\hbox{$\sim$}}}\hbox{$>$}}}}

\newcommand{\beq}{\begin{equation}}
\newcommand{\eeq}{\end{equation}}
\def\beqa{\begin{eqnarray}}
\def\eeqa{\end{eqnarray}}
\def\LCDM{\ensuremath{\Lambda}CDM}

\def\head{ \vbox to 0pt{\vss \hbox to 0pt{\hskip 440pt\rm
      LA-UR-10-07069\hss} \vskip 25pt}}
\def \xoff {\ifmmode x_{\rm off} \else $x_{\rm off}$ \fi}
\def \rhorms {\ifmmode \rho_{\rm rms} \else $\rho_{\rm rms}$ \fi}

\def \kms {\ifmmode  \,\rm km\,s^{-1} \else $\,\rm km\,s^{-1}  $ \fi }
\def \kpc {\ifmmode  {\rm kpc}  \else ${\rm  kpc}$ \fi  }  
\def \hkpc {\ifmmode  {h^{-1}\rm kpc}  \else ${h^{-1}\rm kpc}$ \fi  }  
\def \hMpc {\ifmmode  {h^{-1}\rm Mpc}  \else ${h^{-1}\rm Mpc}$ \fi  }  
\def \Mpch {\ifmmode  {h^{-1}\rm Mpc}  \else ${h^{-1}\rm Mpc}$ \fi  }  
\def \Msun {\ifmmode {\rm M}_{\odot} \else ${\rm M}_{\odot}$ \fi} 
\def \hMsun {\ifmmode h^{-1}\,\rm M_{\odot} \else $h^{-1}\,\rm M_{\odot}$ \fi}

\def \LCDM {\ifmmode \Lambda{\rm CDM} \else $\Lambda{\rm CDM}$ \fi}
\def \sig8 {\ifmmode \sigma_8 \else $\sigma_8$ \fi} 
\def \OmegaM {\ifmmode \Omega_{\rm m} \else $\Omega_{\rm m}$ \fi} 
\def \Omegab {\ifmmode \Omega_{\rm b} \else $\Omega_{\rm b}$ \fi} 
\def \OmegaL {\ifmmode \Omega_{\rm \Lambda} \else $\Omega_{\rm \Lambda}$\fi} 
\def \Deltavir {\ifmmode \Delta_{\rm vir} \else $\Delta_{\rm vir}$ \fi}
\def \rhocrit {\ifmmode \rho_{\rm crit} \else $\rho_{\rm crit}$ \fi}
\def \rhou {\ifmmode \rho_{\rm u} \else $\rho_{\rm u}$ \fi}
\def \zc {\ifmmode z_{\rm c} \else $z_{\rm c}$ \fi}

\def\head{
.ps \vbox to 0pt{\vss
                   \hbox to 0pt{\hskip 440pt\rm LA-UR-10-07069\hss}
                  \vskip 25pt}}
\title[Strongly Coupled Cosmologies I] {Strongly Coupled Dark Energy
  Cosmologies\\ yielding large mass Primordial Black Holes }

\author[S.A. Bonometto \etal] { Silvio A.
  Bonometto$^{1,2}$\thanks{E-mail: silvio.bonometto@inaf.it}, Roberto
  Mainini$^3$\thanks{E-mail:roberto.mainini@mib.infn.it}, Marino Mezzetti$^{1,2}$\thanks{E-mail:marino.mezzetti@inaf.it}\\
$^1$ Astronomy Unit, Physics Department, Trieste University,   Via
  Tiepolo 11, I-34143 Trieste, Italy \\ 
$^2$ I.N.A.F., Osservatorio
  Astronomico di Trieste, I-34143 Trieste , Italy\\ 
$^3$ Physics Department G.~Occhialini, Milano--Bicocca University,  
  Piazza della Scienza 3, I-20126 Milano, Italy}

\date{Accepted XXXX . Received XXXX; in original form XXXX}
\pubyear{2018}

\begin{document}
\label{firstpage}
\pagerange{\pageref{firstpage}--\pageref{lastpage}} 
\maketitle

\begin{abstract}
  
Large primordial Black Hole (PBH) formation is enhanced if strongly
coupled scalar and spinor fields ($\Phi$ and $\psi$) are a stable
cosmic component since the primeval radiative expansion (SCDEW
models). In particular, we show that PBH formation is easier
 at a specific time, i.e., when the asymptotic mass $m_H$,
  acquired by the $\psi$ field at the higgs scale, becomes dominant,
so that the typical BH mass $M_{BH}$ depends on $m_H$ value. For
instance, if $m_H \sim 100\, $eV (1 keV) and the coupling $\beta \sim
8.35$ (37), PBH with $M_{BH} \simeq 10^7$--$10^8 M_\odot$ ($\sim
10^3$--$10^4\, M_\odot$) could form. The very mechanism enhancing PBH
formation also causes technical difficulties to evaluate the transfer
function of SCDEW models at high $k$. A tentative solution of this
problem leaves only minor discrepancies from \LCDM, also at these
scales, gradually vanishing for greater $m_H$ values. We conclude
that, for suitable parameter choices, SCDEW models could be the real
physics underlying \LCDM, so overcoming its fine tuning and
coincidence problems, with the extra bonus of yielding large BH seeds.

\end{abstract}

\noindent
\begin{keywords}

  cosmology: dark matter,dark energy--black holes

 \end{keywords}

\section{Introduction} \label{sec:introduction}

The possible relevance of a cosmic scalar field $\Phi$, coupled to
Dark Matter (DM), was envisaged even before SNIa Hubble diagrams
(\cite{riess,perlmutter}) forced us to conform with the existence of
Dark Energy (DE). The option of DE being such self--interacting scalar
field $\Phi$ was then widely explored. As first outlined by
\cite{wetterich}, \cite{amendola2000} and \cite{amendola2002}, it
could ease some \LCDM conundrums, also allowing DE to be a substantial
cosmic component through the whole matter--dominated era.

Their approach however required a self--interaction potential
$V(\Phi)$, including specific parameters. In turn, no initial
conditions for $\Phi$ needed to be specified, when using tracking
potentials (\cite{Steinhardt}, see also \cite{ratrapeebles},
\cite{wetterich} , \cite{SUGRA}).

The \LCDM option however prevailed as, in spite of the extra
parameter(s) in $V(\Phi)$ expressions,
data fitting did not improve. The option of \LCDM being
inadequate to fit (new) data was then explored just by
testing 
a DE state parameter
$w \neq -1$ or
being
a first degree polynomial $w = w_0 + w_a(1-a)$ (\cite{linder}).

A $\Phi$--DM interaction was then discussed, in a fully different
context, by \cite{blvs}. Instead of displaying its main action in
``recent'' times, they supposed it to modify the radiative
era. Uncoupled DM and kinetic--$\Phi$ densities would then scale
$\propto a^{-3}$ and $a^{-6}$.
A suitable energy flow from coupled DM (coDM,
a spinor field $\psi$)
to $\Phi$, however allows for both densities to be $\propto a^{-4}$,
so keeping them a constant fraction of the radiative cosmic content,
and allowing for a primeval Conformally Invariant (CI)
expansion. \cite{blvs} then
verified
that such regime is not only possible, but is a cosmic attractor.
$\Phi$ could be a scalar field involved in inflation while, at low
$z$, it will safely become DE.

Before discussing the successive work on fluctuation evolution, let us
then soon outline the main point made by this paper: without {\it
  ad--hoc} assumptions, this kind of cosmologies allows for the
formation of large mass Primordial Black Holes (PBH), {\it when the
  mass $m_H,$ acquired by the $\psi$ field at the higgs scale, becomes
  dominant}. PBH average mass is given by the
expression
\begin{equation}
  M_{BH} \sim 3.32 \times 10^7 M_\odot \left(100\, {\rm eV}
  \over m_H \right)^{3.31}~,
  \label{mbh}
\end{equation}
when $\beta$ is
is selected to allow for
close values of early wDM (see below) and coDM densities.  Giant BH as
those in high--$z$ QSO's could then form through matter accretion,
without pushing it to the Eddington limit. Moreover, in principle, PBH
number density can approach the observed galaxy number, while the late
PBH masses depend on individual histories of matter accretion. This
however requires BH seeds with masses $ \ll
10^7 M_\odot$ (as preferred $m_H$ range, values $\cal O$$(1\, $keV)
could be suggested), but immediately rises the question of early
cosmic reionization and microlensing. We shall further comment on
this point later on, but the bulk of this discussion is postponed to
further analysis.

Let us now summarize further previous work on these models: (i) In
\cite{bm1} fluctuation modes were studied and an algorithm yielding
linear transfer functions was described. (ii) In \cite{bmm} the
lagrangian approach was deepened, and a late interaction screening,
due to the coDM field $\psi$ acquiring a mass at the higgs scale, was
also envisaged. This is the very mechanism exploited here. (iii) In
\cite{maccio2015}, N--body simulations of this cosmology, named SCDEW
(Strongly Coupled Dark Energy plus Warm DM), were discussed. They
showed that, when compared with sub--galactic data (MW and M31
satellites, dwarf rotation curves, etc.), SCDEW performed much better
than \LCDM N--body simulations. (iv) A series of papers
[\cite{bm1,bmm,bm2,bm3,bm4}] was then dedicated to analyse the
peculiar evolution of coDM fluctuations, through the radiation
dominated era: In spite of radiation domination, their amplitude
fastly increases.
The numerical algorithm built by \cite{bm1} allows us to follow
  its growth during horizon crossing, when it is linear but fully
  relativistic, as well as later on, until the amplitude allows a
  linear treatment. If we describe such growth by a power law
  $\delta_{co} \propto a^\alpha$, we find an exponent $\alpha \geq 2$
  around horizon crossing and until $\delta_{co}$ overcomes its
  horizon value by 1-1.5 o.o.m.; then the rate of increase softens and
  we gradually settle on a regime $\alpha \simeq 1.6$, as already
  envisaged by \cite{amendola2002}. A still faster growth is expected
  when approaching non--linearity; this will occur a suitable time
  after horizon crossing, whose precise evaluation is made possible by
  the fair treatment of horizon crossing. The very non linear growth
  regime was then explored by using
the spherical top--hat approximation. Their findings will be resumed
here below.

It is also worth mentioning soon that the very rapid growth of coDM
perturbations plays a key role in yielding small scale fluctuations of
the warm DM (wDM) component which, in SCDEW cosmologies, is supposed
to dominate the present DM density, although the warm particle mass
$m_w$ is small.
The relation between wDM and coDM will be furtherly discussed below.

Let us outline that a part of the above findings were again outlined
in a recent work by \cite{amenwett}, with no mention of the above
literature. Their basic aim was to outline that the early formation of
coDM non--linearities could favor PBH formation, in an epoch when
neither radiation nor matter fluctuation amplitudes are allowed to
grow. Accordingly, their paper was the first to outline a possible
relation between SCDEW models and PBH formation. However, they focused
on the early reach of a non--linear regime while, according to the
analysis in this work, PBH formation is hardly directly enhanced by
that. On the contrary, a key role is played by the coDM field
acquiring a mass at the higgs scale. This option, not even mentioned
by them, seems to bear a key role in favouring PBH formation on a
specific mass scale range.
Such acquisition, therefore, is not only a ``screening mechanism'',
whose need was mentioned also by \cite{amenwett} to allow for a late
data fitting, but also the key to open a passage through the
virialization wall.

In fact, \cite{bm3,bm4} had showed that spherical top--hat
fluctuations, during the radiation dominated era, meet virialization
conditions at a low density contrast $\Delta_{co,v} \sim 28$ and, even
more significantly, such {\it virialization is just a transitory
  step}, as the peculiar dynamics of coupled particles causes a {\it
  fast dissolution of virializing lumps}.
Lump dissolution had been independently noticed by \cite{2016},
  in a set of numerical simulations involving particles with variable
  mass, although within a different context.

Our point here, however, is that the action of intrinsical forces
pushing towards disruption of spherical geometry as well as the very
post--virialization dissolution are suppressed when coupling fades,
because of higgs screening. This conclusion is obtained by extending
the study of spherical top--hat density enhancements to the epoch when
the acquisition of a (tiny) mass by $\psi$, at the higgs scale, causes
a rapid fading of the $\Phi$--$\psi$ effective coupling $\beta_{eff}$;
this is one of the main technical contributions of this work. In
particular, we find a significant peak on the virial density contrast
$\Delta_{co,v}$, reaching values $ \sim 500$ (or even much more,
depending on $m_H$ value) while $\beta_{eff}$ fades, to later
reconverge to low values. By itself, however, the reach of such larger
$\Delta_{co,v}$ {\it does not} ease PBH formation. We rather argue
that the same physical reasons allowing the reach of an anomalously
large $\Delta_{co,v}$, in a narrow and specific scale interval, can
also favour the prosecution of spherical collapse, towards its
relativistic regime.
In this connection, the question of the actual likelihood of a
spherical geometry, as a function of the fluctuation amplitude at the
horizon, will be suitably debated.

The redshift $z_P$ when $\beta_{eff}$ fades --and, therefore, the mass
scale $M_P$ of the $\Delta_{co,v}$ peak-- are set by the asymptotic
mass $m_H$. Accordingly, the preferred value of PBH is the mass scale
of fluctuations reaching the horizon so earlier to reach the peak at
$z_P$.. This sets the $m_H$ dependence of PBH mass scale outlined in
eq.~(1), and already outlines that the choice of $\beta$ has little
impact on that.
Such impact is even smaller if we keep to the assumption of early
close values for coDM and wDM.

It is however clear that the geometry of most fluctuations is not
spherical. A better approximation to
treat their initial non--linear stages
could be based on the Zel'dovich pancake approach. If, as we argue,
dissolution mechanisms display their action soon, when a density
contrast $\sim 3$--4 is approached, this technique also enables us
to provide approximated, but realistic, SCDEW model spectra at large
$k$ values.

 Altogether, SCDEW models are characterized by standard radiation,
  neutrino and baryon components. DE is a quintessential field $\Phi$.
  The option of $\Phi$ having a role in inflation is open, as its
  energy density could keep non--negligible since then, for its growth
  through the intermediate eras is just logarithmic. It is also worth
  outlining soon that, although $\Phi$ self--interaction is bound to
  play a key role, no $V(\Phi)$ potential needs to be specified (but
  see below). A {\it dual} Dark Matter component is then assumed,
  comprising coDM (coupled with $\Phi$) and wDM (uncoupled and
  light). Most of the roles of coDM could also be covered by a scalar
  field $\chi$, an option not deepened here. Rather, we stress the
  option that coDM and wDM are the coupled and uncoupled components of
  the same spinor field. This is suggested by the fact that viable
  models are however characterized by close early densities
  $\rho_{\Phi}$ and $\rho_{co}$.

  Accordingly, we require the masses of coDM and wDM particles to be
  equal, so that $m_H$ is not only the asymptotic mass of coDM,
  acquired at the higgs scale, but also wDM quanta are supposed to
  acquire the same mass. Most model features, however, do not depend
  on the two spinor fields quanta sharing {\it exactly} the same early
  densities, as ratios in an interval close to unity however yield
  viable models.

More in detail, most results of this paper are given for a model where
$m_H = 100\, $eV and $\beta = 8.35$, yielding a (primeval) density
ratio (wDM/coDM) $\simeq 0.9\, $. The values of other cosmological
parameters are: $\Omega_b = 0.049$, $\Omega_{d} = 0.6824$, $h_0 =
0.671$, $T_0 = 2.726$ (symbols keep their usual meanings). As shown by
\cite{maccio2015}, such a low $m_H$ value eases the problems
$\Lambda$CDM N--body simulations exhibit, at scales close or below the
galactic scale. However, for the sake of comparison, another model,
with $m_H = 1\, $keV and $\beta = 37$, yielding an equal density
ratio, is also considered.

The plan of the paper is as follows. In Section 2 we provide a more
detailed reminder on SCDEW models, namely on their background
features.  In Section 3 we discuss spherical density enhancements,
when taking into account both coDM and the other components. A
particular emphasis concerns the approach to redshift values when the
DM--$\Phi$ coupling fades. Section 4 is then devoted to discuss
virialization. Section 5 then shows why $\Delta_{co,v}$ achieves a
maximum value around $\sim 10^7$--$10^8 M_\odot$. The stability of
virialized structures is then debated in Section 6. In Section 7, we
then use the tools illustrated in previous Sections, to treat the
question of average density enhancements of realistic shapes. A
discussion Section then summarizes the paper findings, before drawing
our conclusions in the last Section.

\section{SCDEW cosmologies: a fast reminder}

This Section is devoted to resume the main features of SCDEW models.
In the paragraph below eq.~(8), however, we report a numerical
coincidence, never outlined before; its significance is hard to
estimate.

\subsection{Background dynamics}

Let the background metric read
\begin{equation}
ds^2 = a^2(\tau) (d\tau^2 - d\lambda^2)~,
\label{metric}
\end{equation}
$\tau$ being the conformal time. The state equation of a purely
kinetic scalar field $\Phi$, whose Lagrangian
\begin{equation}
{\cal L}_f \sim \partial^\mu \Phi  \partial_\mu \Phi ~,
\label{freelagrangian}
\end{equation}
yields then $w=p_k/\rho_k \equiv 1$ ($p_k,~\rho_k:$ pressure, energy
density; the suffix $_k$ stands for kinetic, as the above lagrangian
contains no potential term). Accordingly, $\rho_{k} = \dot \Phi^2/2a^2
\propto a^{-6}$.

It is also known that non--relativistic DM density $\rho_{co} \propto
a^{-3}$, its state parameter being $w=0$. A suitable energy flow from
DM to $\Phi$ could speed up DM dilution while $\rho_k$ dilution slows
down, so that both dilute $\propto a^{-4}$.

\cite{blvs} showed this to be consistent with a DM--$\Phi$ coupling
ruled by the equations
\begin{equation}
T^{(\Phi)~\mu}_{~~~\, ~~\nu;\mu} = +C T^{(co)} \Phi_{,\nu}~,
~~~~~~~~~~
T^{(co)~\mu}_{~~~~\nu;\mu} =- C T^{(co)} \Phi_{,\nu}~,
\label{relativisticeq}
\end{equation}
an option introduced since the early papers on DM--DE coupling (see
\cite{amendola2000} and references therein) to hold during the late
expansion stages. In eq.~(\ref{relativisticeq}), here concerning the
early cosmic expansion, $T^{(\Phi,co)}_{\mu\nu}$ are the
stress--energy tensors for the $\Phi$--field, coDM; their traces are
$T^{(\Phi,co)}$; the factor
\begin{equation}
C = b/m_p= (16 \pi/3)^{1/2} \beta/m_p~
\label{couplingconstant}
\end{equation}
($m_p:$ the Planck mass) gauges the DM--DE coupling intensity,
otherwise parametrized by $b$ or $\beta$. In a radiation dominated
epoch, coDM and $\Phi$ densities fall on an attractor.
The existence of such attractor had been already envisaged by
  \cite{amendola2000}, and dubbed ``mode $c_{RM}$'', without any
  further comment; successively, such an attractor was considered in
  connection with a cosmological picture arising from a gravi--dilaton
  string effective action (\cite{amendolaG2002}). Independently of
  these results, \cite{blvs} tested numerically that the attractor is
characterized by (constant) density parameters
\begin{equation}
  \Omega_\Phi = {1 \over 4 \beta^2}~,~~ \Omega_{co} = {1 \over 2 \beta^2}~,
  \label{earlyomegas}
\end{equation}
so that the requirement $\Omega_\Phi+\Omega_{co} \ll 1$ implies that
$\beta \gg \sqrt{3}/2$. Values of $\beta \lesssim 2.5 $ are however
excluded by limits on {\it dark radiation} during BBN or when CMB
spectra form.

If the metric is (\ref{metric}) and lifting the restriction that $\Phi$
is purely kinetic, eqs.~(\ref{relativisticeq}) also read
\begin{equation}
  \dot \Phi_1 + \tilde w{\dot a \over a} \Phi_1 = {1+w \over 2} C a^2 \rho_{co}
  ~,~~~~~~~ \dot \rho_{co} + 3 {\dot a \over a}\rho_{co} = -C \rho_{co} \Phi_1~,
\label{nonrelativisticeq1}
\end{equation}
with $\Phi_1 \equiv d \Phi/d \tau$ and $2 \tilde w = 1+3w - d \ln(1+w)
/ d\ln a$. Eqs.~(\ref{nonrelativisticeq1}) yield $\Phi_1$ and
$\rho_{co}$ evolutions from the $\Phi$--field state parameter $w(a)$,
with no need to specify a $V(\Phi)$ expression.

\cite{bmm} showed eqs.~(\ref{relativisticeq}) or
(\ref{nonrelativisticeq1}) to be consistent with DM being a spinor
field $\psi$, interacting with $\Phi$ through a generalized Yukawa
lagrangian
\begin{equation}
{\cal L}_m = - \mu f(\Phi/m) \bar \psi \psi
\label{interaction}
\end{equation}
provided that
\begin{equation}
\nonumber
f = \exp(-\Phi/m)
\end{equation}
(see also \cite{das}).  Here 2 independent mass scales, $m = m_p/b$
and $\mu = g\, m_p$ are introduced. The constant $b$, however,
coincides with the factor $b$ gauging the DM--$\Phi$ interaction
strength in eq.~(\ref{couplingconstant}), so that $C=1/m$; on the
contrary, $g$ and an additive constant on $\Phi$ keep undetermined.

Let us however outline a numerical coincidence: we can assume $\Phi
\equiv m_p$ at the Planck scale, by taking $g = 2\pi \, e^{-b}$. This
was also done in previous work, finding a fair data fit when $b =
4(\pi/3)^{1/2}\beta \simeq 40$; this $b$ value is also close to  the one mostly
 considered in this paper. It is then noticeable
that, with this choice, $\mu$ is of the order of the electroweak (EW)
scale (more precisely, by forgetting the (arbitrary) factor $2\pi$, we
have: $\ln(m_p/100\, {\rm GeV}) = 39.34$, so that $\beta =
9.61$). But, of course, it is fully licit to forget such coincidence
and explore any other interval. Let us however keep
\begin{equation}
f = \exp[-C(\Phi-\Phi_p)]~~~ {\rm with} ~~~ \Phi_p \equiv m_p~,
\label{fff1}
\end{equation}
with $\mu = 2\pi \, m_p$ in eq.~(\ref{interaction}).  This choice is
quantitatively relevant, when we pass to consider a higgs coupling
screening (\cite{bmm}); however, close values would not yield
significant changes.

Let us then recall that the particle number operator of a spinor field
$n \propto \bar \psi \psi$. Accordingly, the coDM density reads
\begin{equation}
  \rho_{co} = \mu f(C\Phi) \bar \psi \psi
  \label{rhoc}
\end{equation}
(formally $= -{\cal L}_m$).  It is then worth focusing on the term
 \begin{equation} 
  {\delta {\cal L}_m \over \delta \Phi} \equiv [{\cal L}_m]'_\Phi = -
  \mu {f'}_\Phi (C\Phi) \bar \psi \psi = - {{f'}_\Phi (C\Phi) \over
    f(C\Phi)} \rho_{co} = C\rho_{co}
\label{eulerolagrange}
\end{equation}
of the Euler--Lagrange equation which, multiplied by a suitable
factor, stands at the r.h.s. of the first
eq.~(\ref{nonrelativisticeq1}).  Incidentally,
eqs.~(\ref{nonrelativisticeq1}) can be soon integrated, yielding
\begin{equation}
  \Phi_1 = C/\tau~,~~~ \rho_{co} \propto a^{-4}
  \label{integre}
\end{equation}
during the radiative expansion. Of course, also
\begin{equation}
  \rho_k = {\Phi_1^2 /( 2 a^2)}
\end{equation}
then dilutes as $a^{-4}$. This is why $\Omega_\Phi$ and $\Omega_{co}$
keep constant.

\begin{figure}
\vskip -2.truecm
\centering
\includegraphics[width=8.5cm]{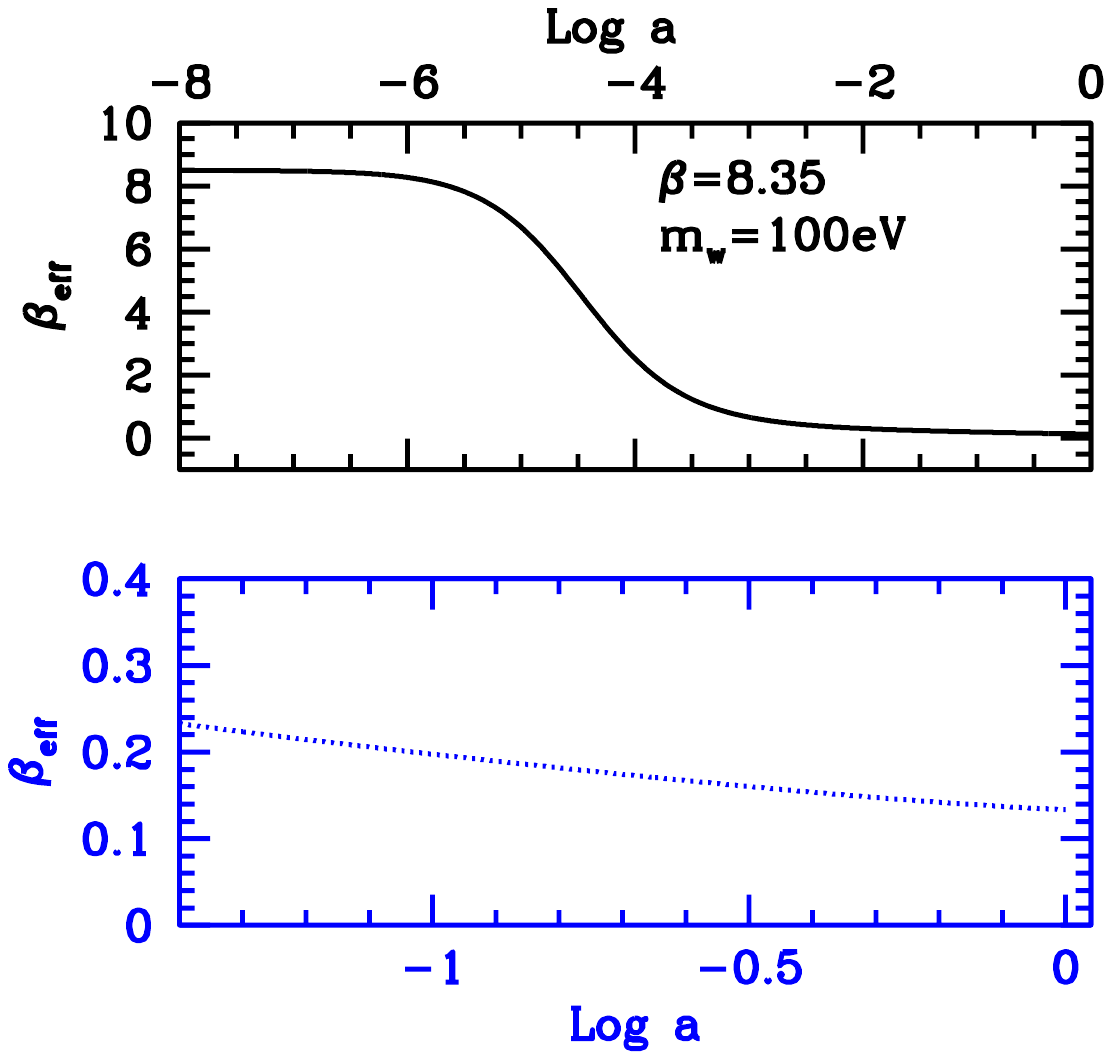}
\vskip -4.truecm
\includegraphics[width=8.5cm]{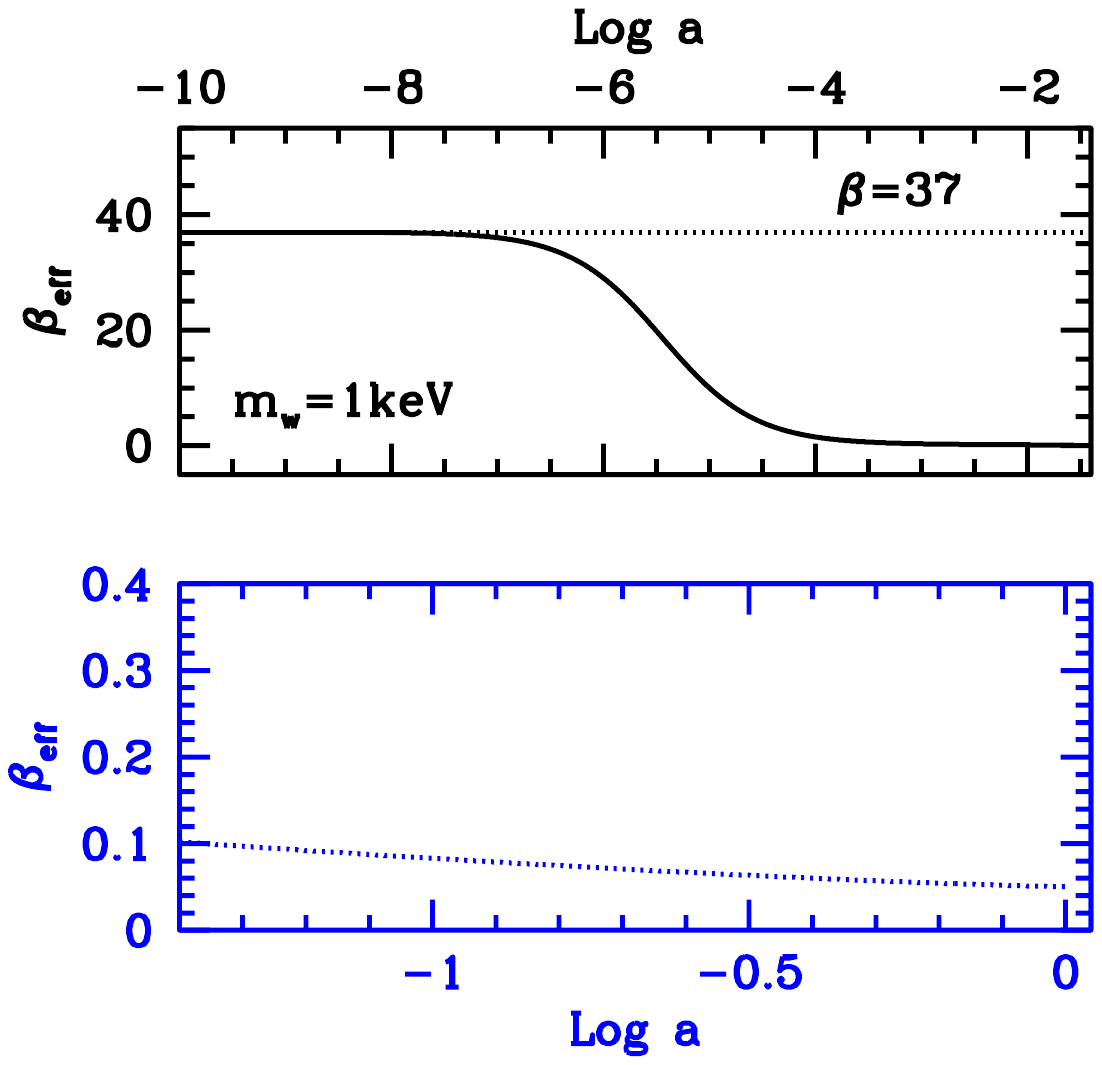}
\vskip -2.3truecm
\caption{As a consequence of $\psi$ acquiring an asymptotic mass at
  the higgs scale, the coupling constant decreases, about the redshift
  when warm DM turns non--relativistic (wDM particle masses in the
  upper figures frames). In the bottom blue plots, we show values
  close to $z=0$. Notice that, due to the longer $\beta_{eff}$
  decrease time, its low--$z$ values of are even smaller than in
  greater $\beta$ case. }
\label{cou10}
\end{figure}   

Let us now consider the effects of asymptotic mass acquisitions. In
fact, at the higgs scale, the lagrangian (\ref{interaction}) shall
become
\begin{equation} 
\label{higgslagrangian}
\tilde {\cal L}_m = -[\mu f(\Phi /m) + \tilde \mu ] \bar \psi \psi
\end{equation}
(so violating primeval CI). Then, by re--doing the functional
differentiation in eq.~(\ref{eulerolagrange}), we obtain
\begin{equation}
{\delta \tilde{\cal L}_m \over \delta \Phi} =-\frac{f'(C\Phi
  )}{f(C\Phi)+{\tilde \mu/\mu}}\rho_{co} = {C \over 1 + {\cal R}
  \exp[C(\Phi- \Phi_p)]} \rho_{co} ~.
\label{variablecoupling}
\end{equation}
Here $ {\cal R} = \tilde \mu/\mu $. Accordingly, the dynamical
equations, in the presence of the mass acquired at the higgs scale,
keep the form (\ref{nonrelativisticeq1}), once we replace
\begin{equation}
\nonumber
C \to C_{eff} = {C \over 1 + {\cal R} \exp[C(\Phi- \Phi_p)]}
\end{equation}
and/or
\begin{equation}
   \beta \to \beta_{eff} = {\beta \over 1 + {\cal R} \exp[C(\Phi-
       \Phi_p)]}~.
\label{newC}
\end{equation}
Then, when the $\Phi$ increase causes $\Phi-\Phi_p$ to approach
$-\ln({\cal R})/C$, the denominators in eq.~(\ref{newC}) suppress the
effective coupling intensity. For the sake of example, by assuming
$\beta=10$ (37) and $\tilde \mu = 115\, $eV (1$\, $keV), the
dependence of $\beta_{eff}$ on $a$ is shown in Figure 1 (Figure 2).

Besides of radiation, baryons, and the coupled $\Phi$ (DE) and $\psi$
(coDM) fields, SCDEW models also include a further DM component, that
we shall indicate as wDM (warm DM). In the models considered here, wDM
is assumed to be closely related to coDM, acquiring the same mass
($m_{co}=m_w=m_H$) at the higgs scale
and exhibiting a close early density (in the cases considered,
$\rho_{co}/\rho_{w} \simeq 0.9$). Then coDM has a later rise, as the
flow of energy from it to $\Phi$, yielding $\rho_{co} \propto a^{-4}$
at large $z$, is (almost) cut off only when $\beta_{eff}$ has reached
its low value domain.

The assumption that coDM and wDM have similar masses and early
densities, during the primeval CI expansion, is somehow arbitrary. The
only compulsory requirement on masses and $\beta$ is that they allow
for densities $\rho_i$, at $z=0$ (the index $_i$ labels cosmic
components) fitting observations. Our requirement however reflects the
conjecture that coDM and wDM particles are, somehow, the
$\beta$--coupled and the neutral state of the same particle.  A
priori, one could state that SCDEW models require two DM components.
However, they are safely viable under the above restrictions, as
though we were dealing with a single DM component with 2 charge
states.

When approaching our epoch, we expect a $\Phi$ transition from kinetic
to potential. Rather than dealing with hardly testable $V(\Phi)$
expressions, we model the $w$ transition from +1 to -1, by requiring
\begin{equation}
w(a) = {1 - A \over 1 + A} ~~~{\rm with} ~~~ A = \left(a \over a_{kp}
\right)^\epsilon~.
\label{kp}
\end{equation}
As expected, results are scarcely dependent on the exponent
$\epsilon$, whose arbitrariness somehow mimics the arbitrariness in
the potential choice; $a_{kp} = (1+z_{kp})^{-1}$ is fixed so to obtain
the observational amount of today's DE.

\begin{figure}
\vskip -2.truecm
\centering
\hskip +2.truecm
\includegraphics[width=7.5cm]{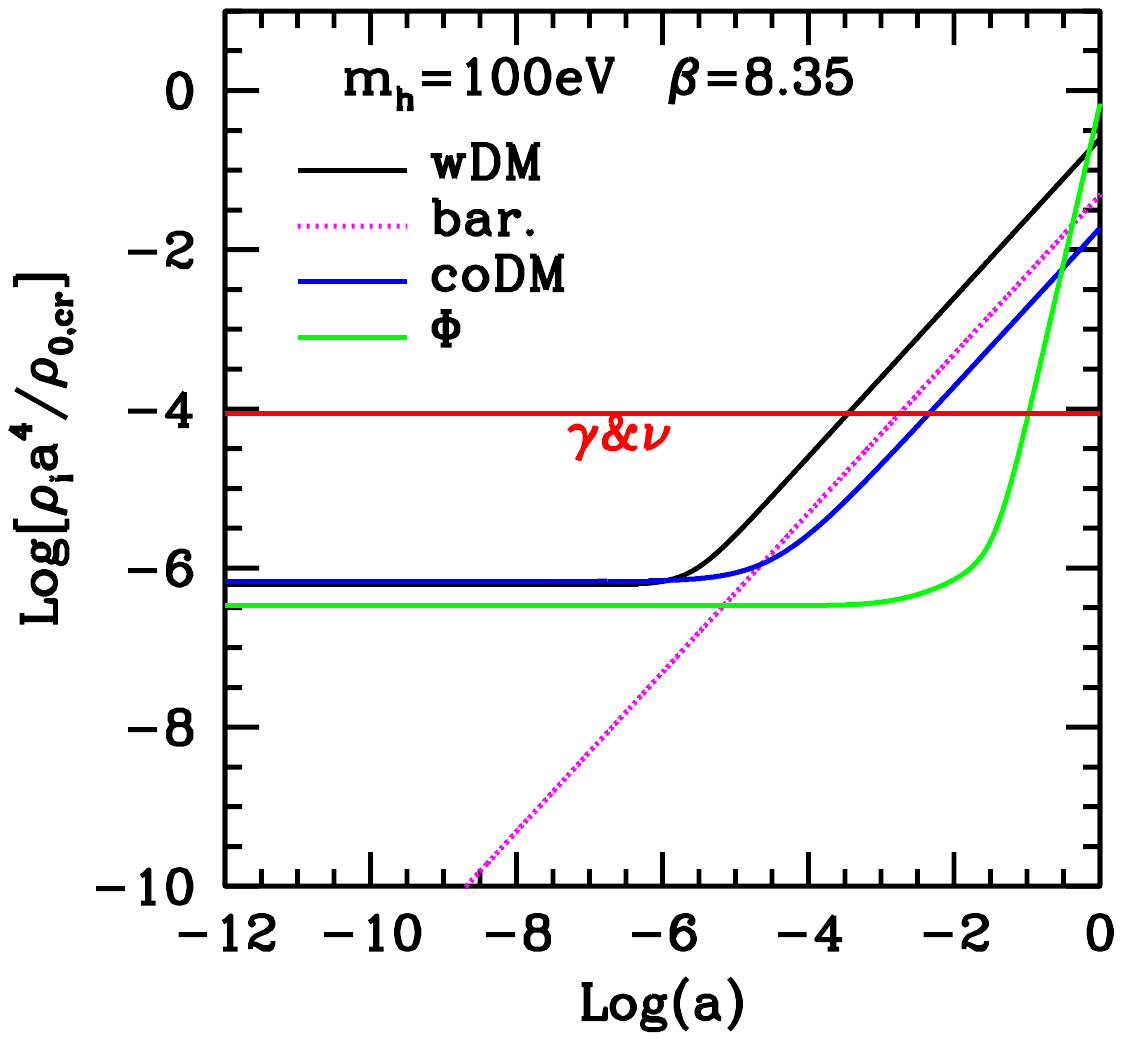}
\vskip -3.49truecm
\includegraphics[width=7.5cm]{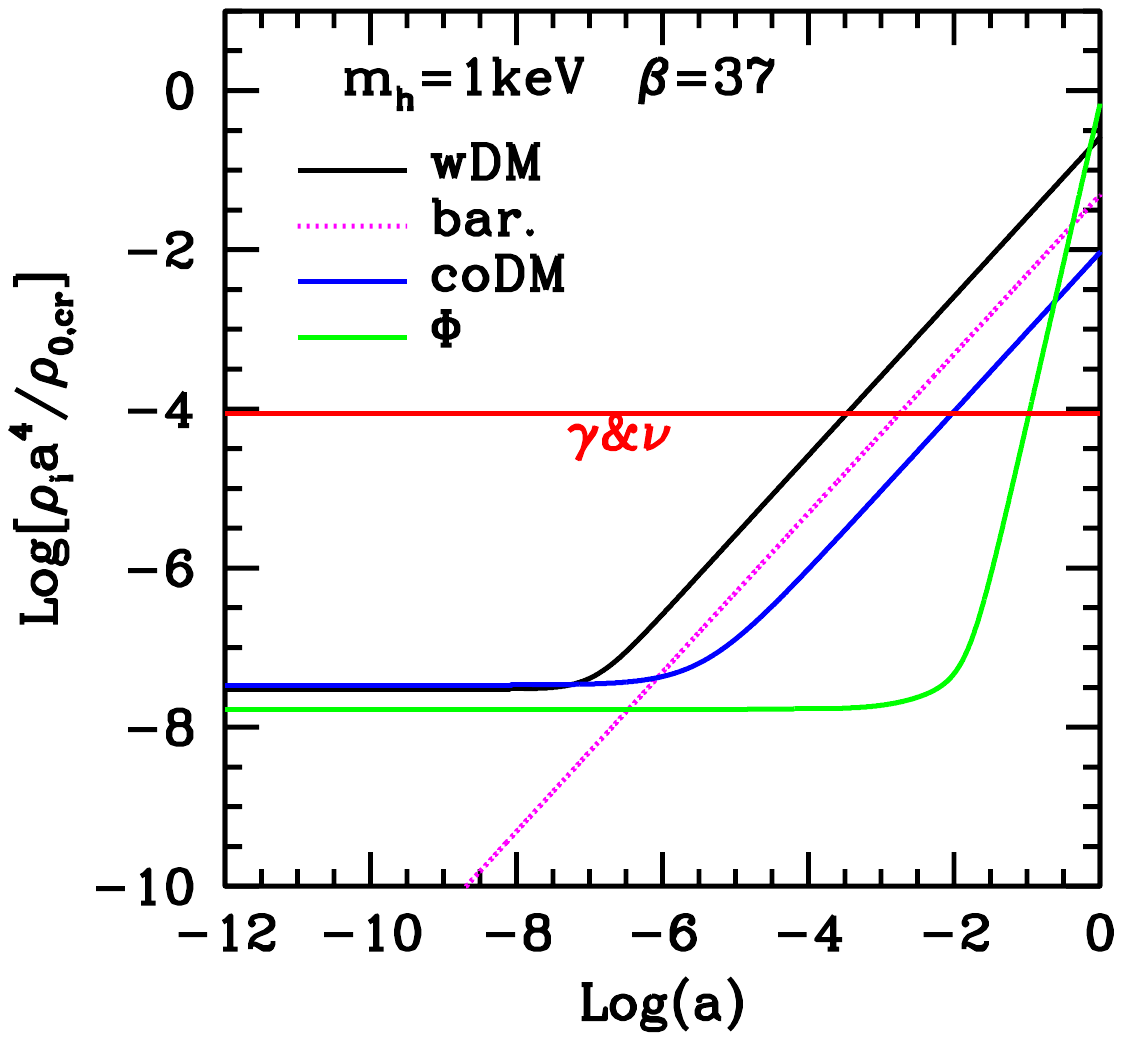}
\vskip -2.1truecm
\caption{Evolution of densities ($\rho_i$) in the 2 cosmologies
  discussed in detail in this work.}
\vskip -.5truecm
\label{densita}
\end{figure}

In Figure \ref{densita} the scale dependence of the densities is
plotted.
Here, DE evolution is drawn by taking $\epsilon=2.9$. Taking different
$\epsilon$ values cannot modify either $\rho_{0,\Phi}$ or
$\rho_{\Phi}$ at high $z$, as the former value is assumed while, at
high $z$, $\Omega_\Phi \equiv 1/4\beta^2$. Therefore, only the
detailed scale dependence, close to $a_{kp}$, is slightly modified.
As a change of $\epsilon$ mimics a change of $V(\Phi)$ expression,
this is a further indication of the serious difficulty that will be
however found to detect $V(\Phi)$ from observational data.

\subsection{Linear fluctuation evolution}

Linear fluctuations in SCDEW models were first discussed in
\cite{bm1}. In a synchronous gauge, the metric shall then read
\begin{equation}
  \label{newmetric}
ds^2 = a^2(\tau) [d\tau^2 - (\delta_{ij}+h_{ij}) dx_i dx_j)]~,
\end{equation}
gravity perturbations being described by the 3--tensor $h_{ij}$, whose
trace is $h$. Besides of density perturbations for all components, one
shall then consider also field perturbations, by assuming
\begin{equation}
\phi = \Phi + {b \over m_p} \varphi~
\label{varphi}
\end{equation}
to be the sum of the background field $\Phi$ considered in the
previous subsection and a perturbation described by $\varphi$.

The whole discussion has many technical aspects that are deepened in
\cite{bm1}. The main critical issue, when trying to use equations
provided in the literature, arises because we set $w(a)$, instead of
the potential.  This is obtainable by replacing
\begin{align}
\nonumber
2V'' =& {A \over 1+A} \bigg\{ {\dot a \over a} {\epsilon \over 1+A}
\left[ (\epsilon-6) {\dot a \over a^3} + 2C {\rho_{co} \over \dot \Phi}
  \right] + \\
 &  + \left[ {\dot a \over a^3} {\ddot \Phi \over \dot \Phi} + {d \over
      d\tau} \left( \dot a \over a^3 \right) \right] \epsilon_6 + 2C
       {\dot \rho_{co} \over \dot \Phi} \bigg\} 
\end{align}
with $A$ and $\epsilon$ defined as in eq.~(\ref{kp}). Here $\rho_{co}$
is the background density of coDM.

\section{Non linearities in the early Universe}

When entering the horizon, in the early Universe, coDM density
fluctuations $\delta_{co,hor} > 0$ exhibit an average amplitude $\bar
\delta_{co,hor}^+ \sim 10^{-5}$, close to the top likelihood
value for positive fluctuations, (supposedly) with a Gaussian
distribution. Fluctuations with a greater horizon amplitude $ F \times
\bar \delta_{co,hor}^+ $ will be also considered below.

\begin{figure}{}
\vskip -3.2truecm
\centering
\includegraphics[width=8.5cm]{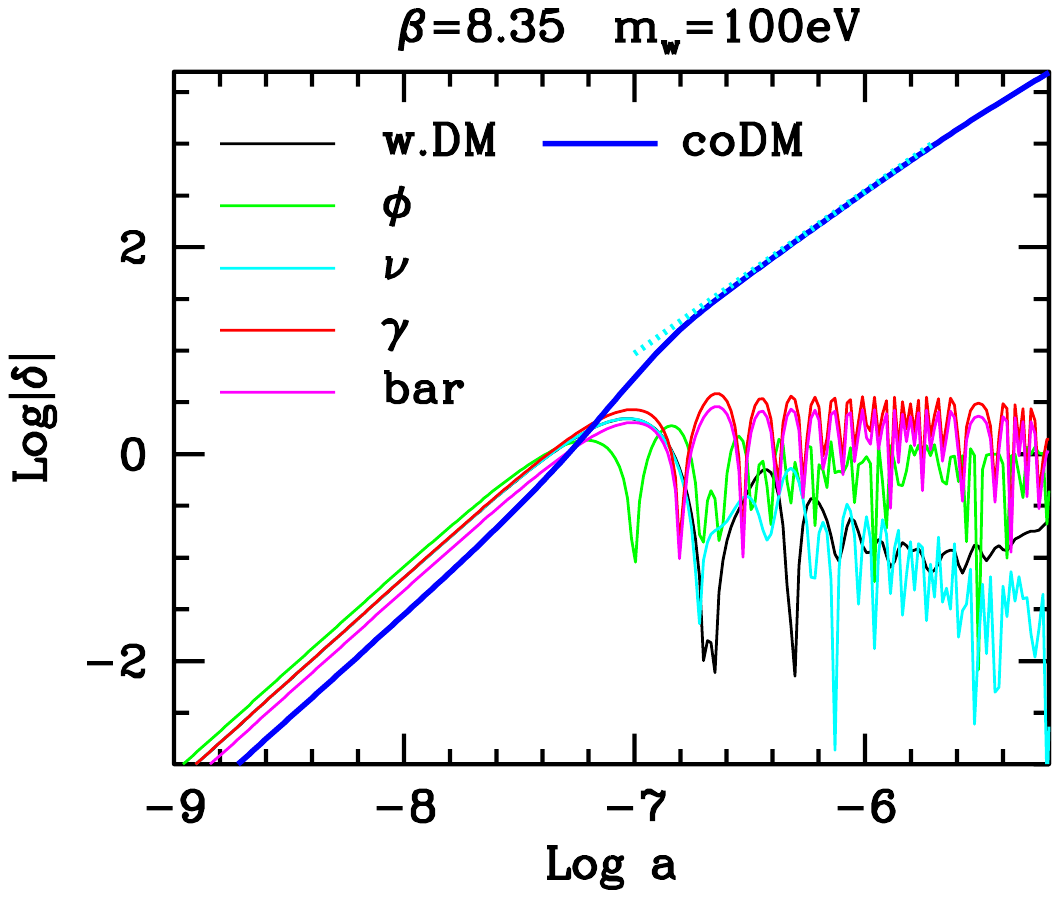}
\vskip -2.7truecm
\caption{Fluctuation evolution in the cosmic components at their entry
  in the horizon (at high $z$). The dotted line has a steepness
  $\alpha=1.6$. }
\label{flevo}
\end{figure}

\begin{figure}{}
\vskip -3.2truecm
\centering
\includegraphics[width=8.5cm]{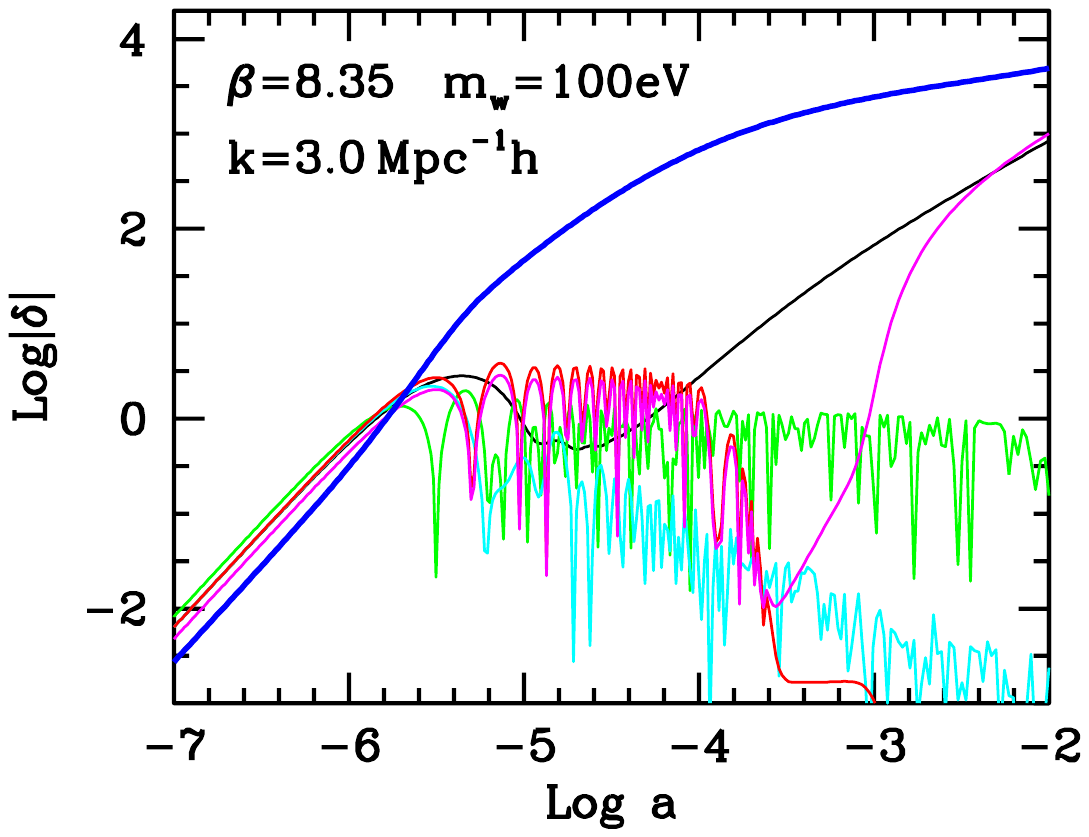}
\vskip -2.7truecm
\caption{Fluctuation evolution in the cosmic components from their
  entry in the horizon until $z \simeq 100$, for model and scale
  indicated in the frame. Model parameters are selected so to cause an
  early coupled--DM non--linearity on such scale.  Colors as in
  previous Figure. }
\label{glevo}
\end{figure}

As shown in Figures \ref{flevo} and \ref{glevo}, obtained from the
linear program by \cite{bm1}, in a synchronous gauge $\delta_{co}$
undergoes an uninterrupted growth, with a greater rate around horizon
crossing. Then, when attaining the non--relativistic regime,
$\delta_{co} \propto a^\alpha$ with $\alpha \simeq 1.6\, .$ It is so
in spite of coDM being $\sim 1\, \%$ (or less) of the total density,
in the radiation dominated epoch.  This behavior is further
illustrated in Figure \ref{glevo}, where we extend the plot of
fluctuation evolution, so to approach $z=0$, for all cosmic
components. All that can be straightforwardly understood, on the basis
of the newtonian limit of coDM dynamics, as discussed by
\cite{maccio2004} and \cite{baldi}, and resumed here below.

Such early growth is critical to allow for SCDEW spectra approaching
\LCDM, up to $k \sim 1$. At greater $k$'s, however, its effects appear
excessive as, even for $\delta_{co,hor} \simeq \bar
\delta_{co,hor}^+$, values $\delta_{co} >\sim 1$ are attained earlier
than ``today''. When this occurs, the linear program yields unphysical
outputs. One of the aims of this work is to show how to deal with such
non--linearities.

The peculiar behavior of $\delta_{co}$, in the non--relativistic
regime, can be understood if taking into account that coupling effects
are then equivalent to: (i) An increase of the effective gravitational
push acting between coDM particles, for the density fraction exceeding
average (while any other gravitational action remains normal). The
increased gravitation occurs as though $G = 1/m_p^2$ becomes
\begin{equation}
G^* = \gamma\, G ~~~~{\rm with} ~~~~ \gamma = 1 + 4\beta^2/3
\label{gstar}
\end{equation}
  
\noindent
(ii) As already outlined in eqs.~(\ref{rhoc})
and (\ref{integre}), coupled--DM particle masses progressively
decline. This occurs while the second principle of dynamics still
requires that ${\bf f} = {\bf p}'$ (here the prime indicates
differentiation in respect to the ordinary time $t$). This yields the
dynamical equation
\begin{equation}
{d{\bf v} \over d\, t} = {{\bf f} \over m_{eff}} + \left|m_{eff}' \over
m_{eff} \right| {\bf v}~,
\label{fuma}
\end{equation}
i.e. an {\it extra--push} to particle velocities, adding to the
external force~{\bf f}. Once eqs.~(\ref{gstar}) and (\ref{fuma}) are
applied, the whole effects of coupling are taken into account.

The self--gravitational push due to $\delta_{co}$ is then proportional to
\begin{equation}
G^* \delta_{co} \rho_{co} = G\, \delta_{co} \rho_{cr} {1 \over 2 \beta^2} \times
\left(1 + {4\beta^2 \over 3} \right) = G\, \delta_{co} \rho_{cr} 
 \left( {2 \over 3} + {1 \over 2\beta^2} \right),
\end{equation}
as though concerning the whole critical density $\rho_{cr}$, although
with an amplitude reduced by a factor (slightly exceeding) 2/3~.  We
must add to that the {\it extra push} due to particle mass decline.

Such fast increase eventually leads $\delta_{co}$ into a non--linear
regime. As a first step, to gain an insight on non--linear evolution,
we can assume $\delta_{co}$ to be the amplitude of a spherical
top--hat density enhancements. Real fluctuations approaching
sphericity are surely rare, at least for $F \sim 1$, although becoming
more likely as $F$ increases (see below). There are however
significant conclusions we can draw from spherical dynamics, that we
discuss in the next Section. There, we prescind from the actual $F$
value, provided that non--linearity is attained when the relativistic
regime, due to horizon crossing, is over.

\subsection{Spherical top--hat dynamics}

Let then $R = ca$ be the spherical top--hat radius ($c:$ comoving
top--hat radius). The relation between $c = R/a$ and the density
contrast $\Delta_{co} =1 + \delta_{co}$ then reads
\begin{equation}
\Delta_{co} = 1+\delta_{co} = \tilde\Delta_{co} \tilde c^3/c^3~,
\end{equation}
$\tilde c$ being the radius at a suitable reference time $\tilde
\tau$; accordingly, by assuming $\delta_{co} \propto \tau^\alpha$,
\begin{equation}
{\dot c \over \tilde c}_{\tilde \tau} = -{\alpha \over 3} {\tilde
  \delta_{co} \over \tilde \Delta_{co}} {1 \over \tilde \tau}
\label{dotc}~;
\end{equation}
this relation allows us to choose, during the linear stages, an
arbitrary time $\bar \tau$ when we start to use $c$ instead of
$\delta_{co}$ to follow the top--hat dynamics, independently of the
value of $F$ at the horizon.

\begin{figure}{}
\vskip -2.truecm
\centering
\includegraphics[width=7.5cm]{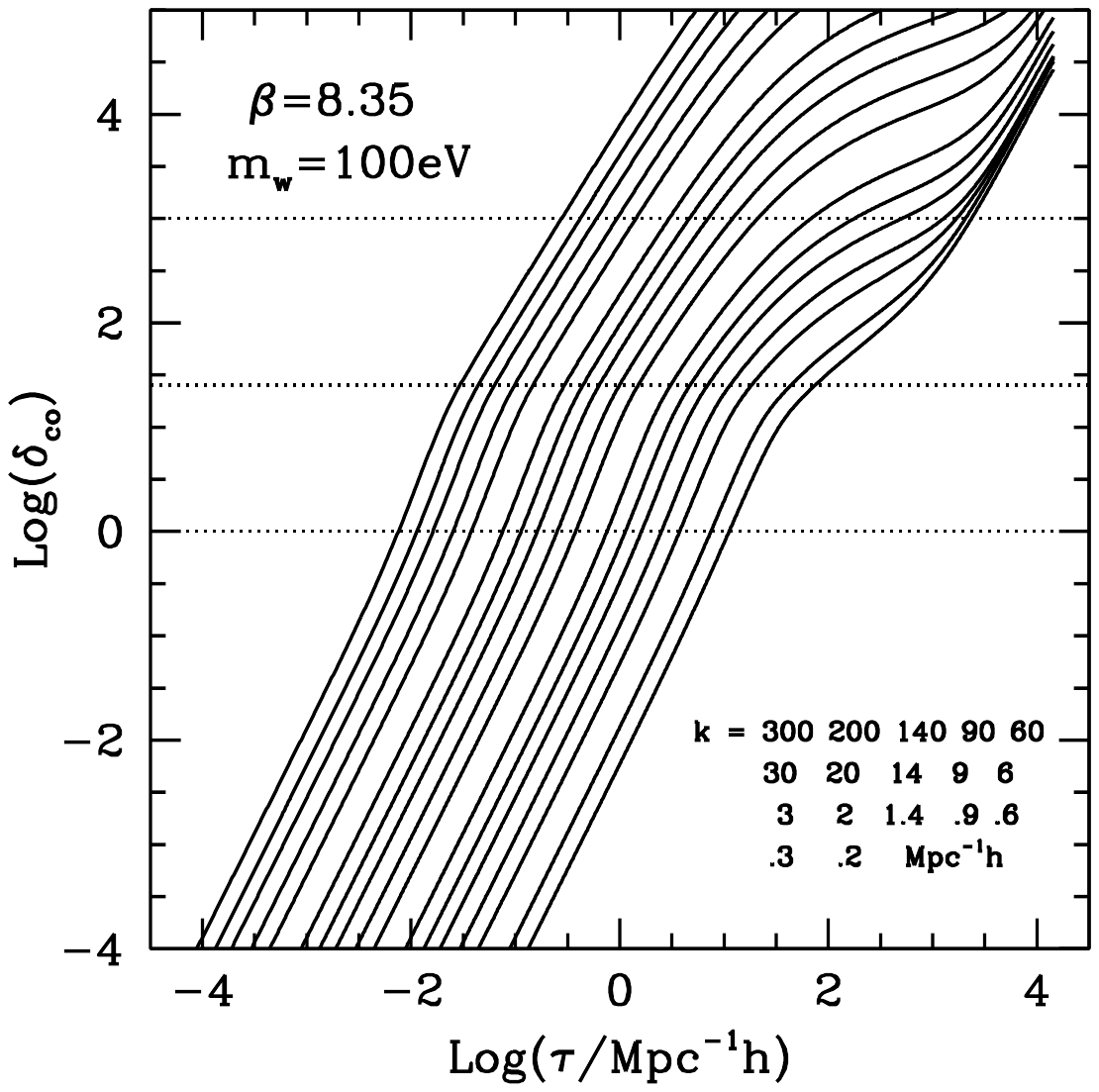}
\vskip -1.7truecm
\caption{$\delta_{co} $ evolution across the horizon and until the reach
  of non--linearity (for average amplitude fluctuations).  At the
  bottom right, the $k$ values considered are listed.  }
\label{many}
\end{figure}

It is however important to set a fair value of $\alpha$, in
eq.~(\ref{dotc}). This need is evident also from Figure \ref{many},
showing the growth rate variation, in the linear regime, for different
$k$ scales.
Let us outline that the numerical algorithm sets to unity the coDM
fluctuation at the horizon entry. In the Figure, non linearity is
assumed to be reached when $\delta_{co} \sim 10^5$. The $k$ values
considered are listed in the bottom right of the Figure. Horizontal
dashed lines indicate the end of the relativistic regime associated to
horizon crossing and the reach of $\delta_{co} = 10^{-2}$ for average
fluctuations. The $\alpha$ dependence on scale is therefore
significant. If greater horizon amplitudes are considered, this latter
line should be lowered and, more importantly, we ought to change the
$\alpha$ value associated to each $k$ scale.


In a previous paper, we dealt with $c$ evolution, by assuming
$\Delta_{co}$ not to cause other component inhomogenities. This may be
a reasonable assumption during the very early expansion, but needs to
be tested when we approach $\beta_{eff}$ fading and matter--radiation
equality~$z_{eq}$.

As some of the key issues of this work arise from the analysis of such
period, we need to deepen the question of other component involvement,
and this point is one of the main technical contributions of this
work.

A critical issue, however, is that a top--hat configuration, involving
relativistic components, would be rapidly smoothed by particle
velocities. If the warm DM mass $m_w \sim 100\, $eV, derelativization
occurs around $z_{eq}$ and, even afterwards (or for greater $m_w$),
particle motions keep non negligible. Henceforth, a test on the
effects of/on other components would be intricate, if we strictly keep
to the model.

The possible impact of other components on top--hat dynamics is
however strengthened, if we assume all of them to be safely
non--relativistic. In this way we can compare the growth obtainable if
neglecting other component gravity, {\it vs.}~an over--modified
growth, with changes exceeding those possibly due to the gravity of
fluctuations in other physical components. On the contrary, the
simultaneous growth of fluctuations in the artificially
non--relativistic component is overestimated and looses much of its
significance.

More in detail: when setting the initial conditions for $c$ evolution,
a sphere of background materials with (comoving) radius $b = c$
overlapping the top--hat, starts to be affected. Initial conditions
for $b$ evolution, in principle, are obtainable in analogy to
eq.~(\ref{dotc}), from the actual values of $\delta_w$, for wDM, and
$\delta_b$, for baryons. Quite in general, $\dot b$ values obtained in
this way exhibit a variable sign with zero average, however being $\ll
\dot \delta_{co}$. Accordingly, we shall simply assume $\dot b=0$.

The main technical problem arises because $c$ grows faster than $b$,
so that the background material sphere acting on $c$ gradually
shrinks. This is why it is convenient to share $b$ in $N$ parts and
consider (sub--)spheres of radii $b_n = b(n/N)$ ($n=1,2,...,N$;
hence $b \equiv b_N.$) When $c$ decreases, it soon becomes
smaller than $b_N$ and then, eventually, smaller than various $b_n$
with $n<N$. Only background materials spheres with $b_{n-1}<c$ are
then possibly acting on $c$ evolution. Furthermore, at most times, it
will be $b_n > c > b_{n-1}$, and interpolation is needed to gauge the
action of a fraction of the $n$--th shell on the coDM sphere radius
$c$.

\begin{figure}{}
\vskip -2.truecm
\centering
\includegraphics[width=8.5cm]{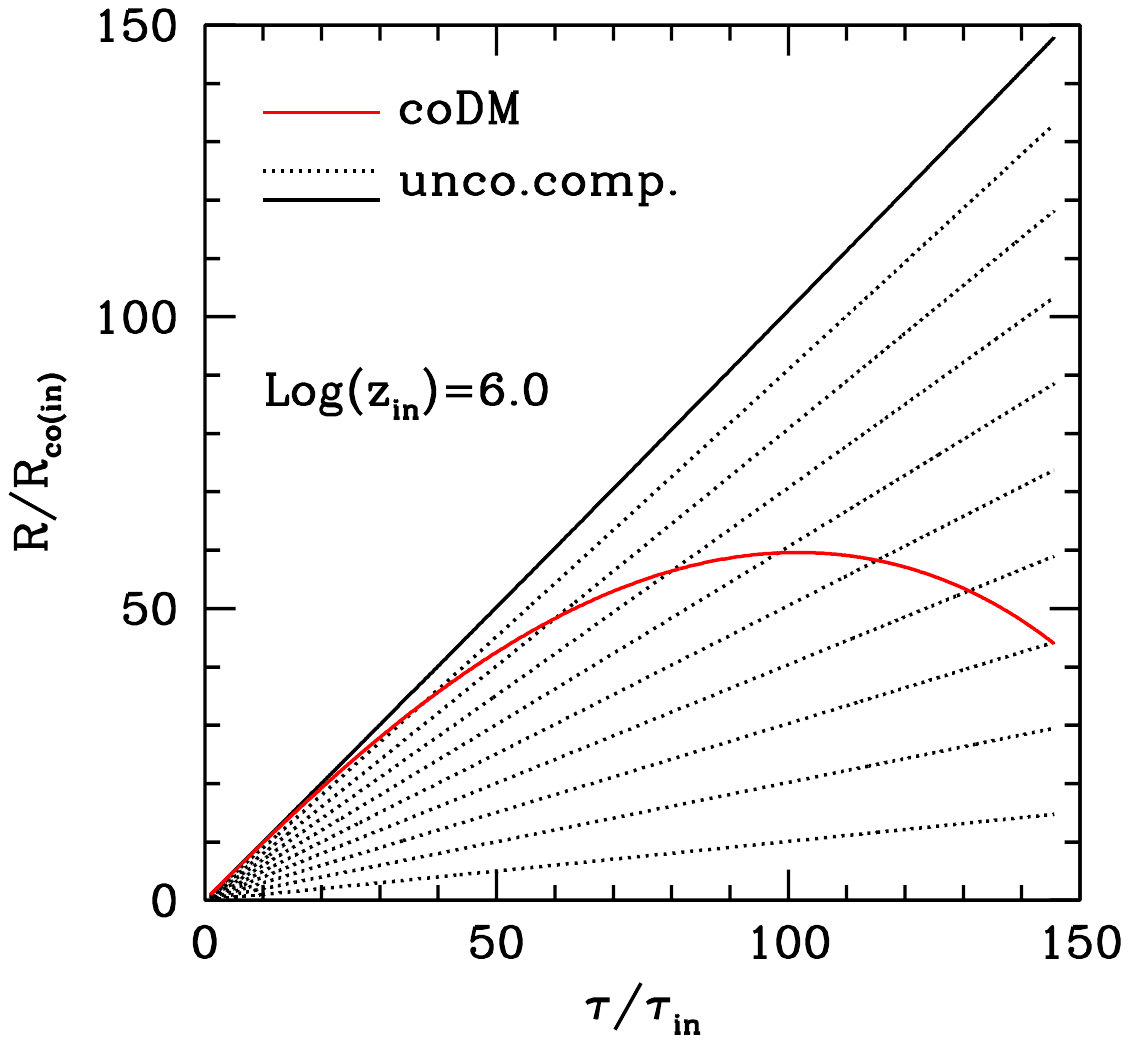}
\vskip -2.7truecm
\caption{Evolution of comoving radii $R=ca$ and $R_n=b_na$ until coDM
  virialization. For graphical reasons, the initial time is indicated
  by a suffix $_{in}$, instead of ~$\tilde {}\, $, as in the text.}
\vskip -.7truecm
\label{layers}
\end{figure}   

\begin{figure}{}
\vskip -2.truecm
\centering
\includegraphics[width=8.5cm]{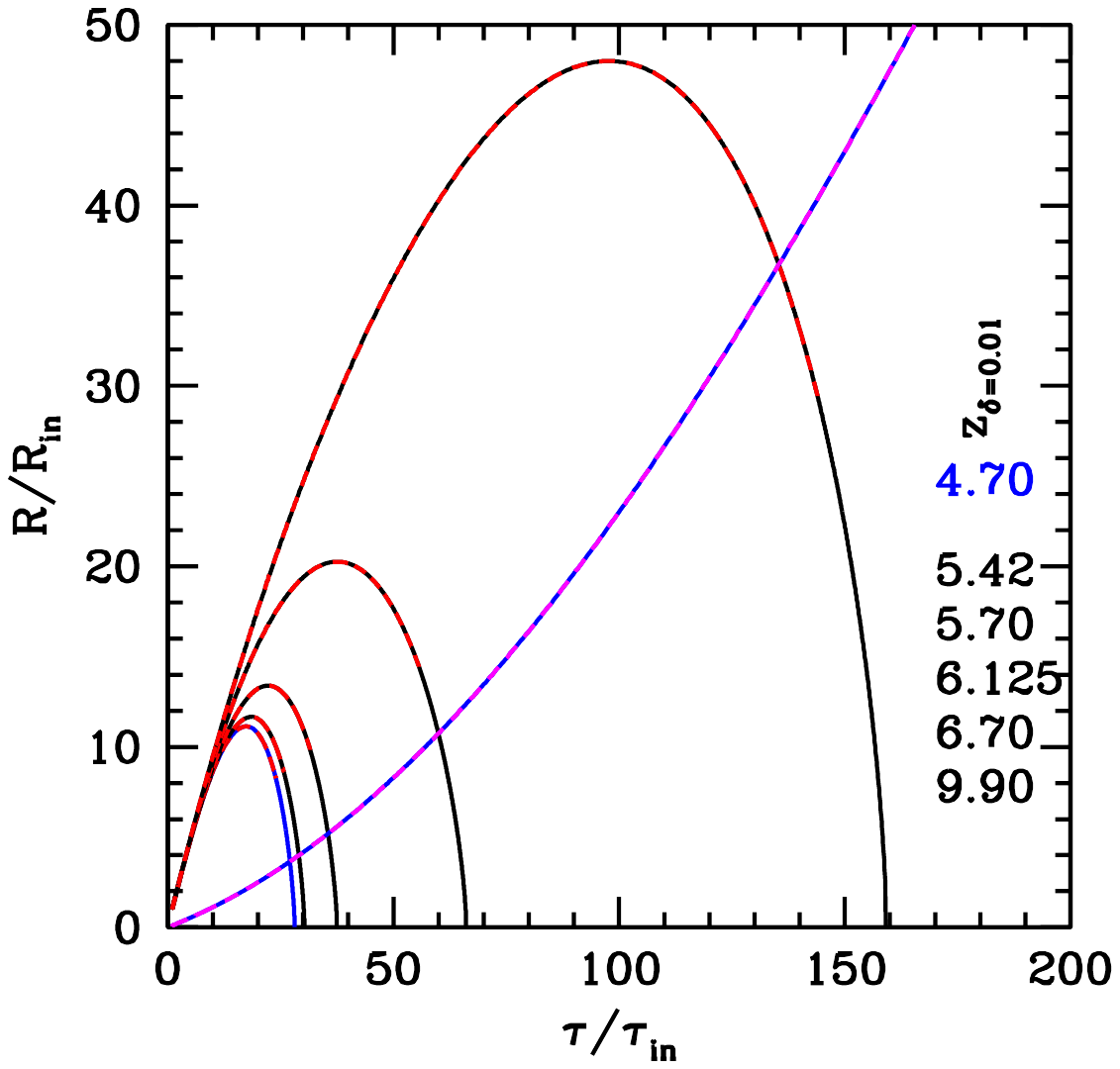}
\vskip -2.truecm
\caption{Evolution of sphere radii, when different values of
  $z_{\delta=0.01}$ are considered. At the low r.h.s. of the Figure
  their logarithms are listed. The evolution becomes increasingly long
  when smaller $z_{\delta=0.01}$ are considered (see also text). For
  the last value (4.70) we plot (in blue) $R/10$ instead of $R$.  In
  top of each curve, the red (magenta) dashed line indicates the
  evolution until virialization conditions are attained.  For
  graphical reasons, as in the previous Figure, the initial time is
  indicated by a suffix $_{in}$, instead of ~$\tilde {}\, $. }
\label{evosphe}
\end{figure}   
\begin{figure}{}
\vskip -2.truecm
\centering
\includegraphics[width=8.cm]{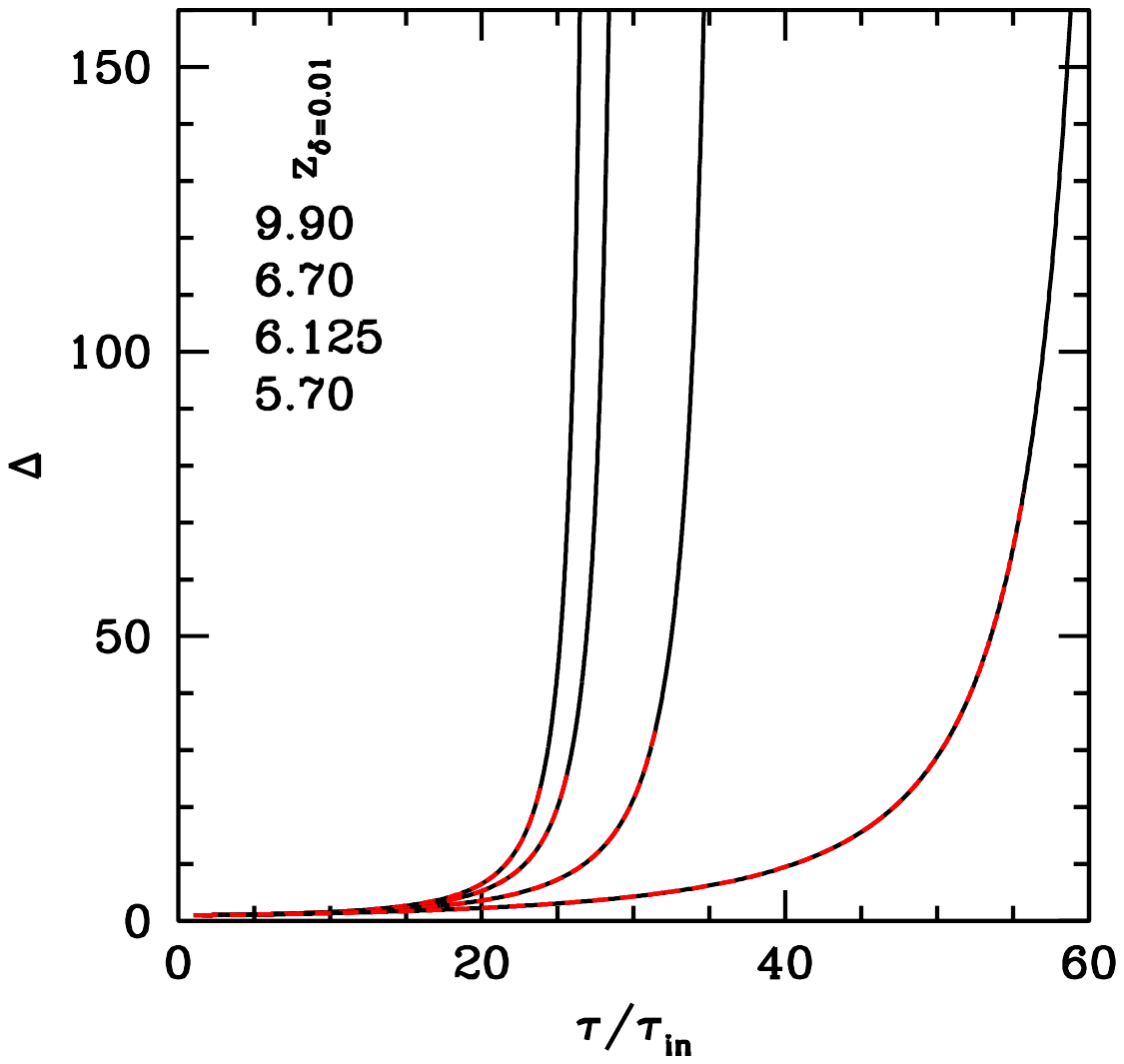}
\vskip -2.3truecm
\caption{Time dependence of $\Delta_{co}$ growth, for the 4 greater
  $z_{\delta=0.01}$ considered in the previous Figure. The red dashed
  part of the curves indicates the growth up to virialization.  At the
  top left, the logarithms of the $z_{\delta=0.01}$ considered.}
\vskip -.35truecm
\label{DCgrow}
\end{figure}   
This technical problem is strictly analogous to the one faced by
\cite{mainini2005} and \cite{mabo}, and is debated in Appendix A. Here,
let us just outline that the actual variables used in dynamical
equations are
\begin{equation}
  x = c/\tilde c, ~~ y_n = b_n/\tilde c~,
\end{equation}
while the independent time variable will also be normalized at the
initial time, by setting $u = \tau/\tilde \tau$.

In the early CI expansion, results are independent from $\tilde \tau$.
Here, however, we extend the treatment to low $z$ values, our main
results concerning times when $\beta_{eff}$ fades. Although dynamical
equations are then formally $\tilde \tau$ independent, several
coefficients enclosed in them exhibit a specific dependence on $\tilde
\tau.$

In principle, results are more and more reliable when greater $N$
values are considered, so that outputs do not rely on the
interpolation inside the $n$--the shell. Our tests however show that
results are already stable for $N=10$, as in Figure \ref{layers}.
Clearly, the deviations of $b_n$ from straight lines are just
marginally appreciable.

This Figure, as well as the whole results on spherical top--hat
evolution, are obtained for a model with $\beta=8.35$ and $m_H= 100\,
$eV. Results will be later extrapolated to other couplings and
asymptotic mass values.

In Figure \ref{evosphe} we then show the evolution of top--hat radii,
and its dependence on the ``initial redshifts'' $z_{\delta=0.01}$,
selected so that the linear growth yields then a coDM fluctuation
amplitude $\delta_{co}=0.01$~. As expected, $R/\tilde R$, starting
from unity, reaches a maximum value, and then re--decreases (for
graphical reasons, in the Figures, $\tilde R$, $\tilde \tau$, etc.,
are replaced by $R_{in},~\tau_{in}$, etc.). The decrease, if we assume
a never--violated spherical symmetry, stops only when re--approaching
$R = 0$ and $\dot R$ yields a velocity approaching the speed of light;
then, non--relativistic equations fail to work.

In top of each black curve a dashed red curve is also plot, stopping
when the virialization conditions are attained. The reach of such
conditions and the significance of red curves are discussed in the
next sub--section.

For the model considered, Figure \ref{evosphe} confirms that, at large
$z$, the growth and recollapse process is substantially independent
from the initial redshift. A small, initial deviation from such
self--similarity takes place when $\log(z_{\delta=0.01}) <\sim 7$,
being visible in our plot for 6.70~. Deviations then become wider; for
$\log(z_{\delta=0.01})= 6.13$, e.g., a significant enhancement of $R$
evolution is appreciable. Further examples, down to
$\log(z_{\delta=0.01})= 5.70$ and 5.42 then show further abrupt
enhancements, with a duration of the whole process finally boosted by
a factor $\sim 30~.$ This is however quite small, in comparison with
the case $\log(z_{\delta=0.01}) = 4.70$ or still smaller initial
redshift values. This last example is shown in the Figure by plotting
$(R/\tilde R)/10$. In spite of that, the Figure frame is unsuitable to
contain the whole evolution, leading to a top $R/\tilde R \simeq 1272$
when $\tau/\tilde \tau \simeq 369$, while the recollapse to zero takes
place when $\tau/\tilde \tau \simeq 476\, $, i.e. at $z \sim 50~.$ For
just slightly smaller $z_{\delta=0.01}$ values, full recollapse is not
yet attained at the present time.

The evolution of the density contrast $\Delta_{co}$ is then strictly
related to $R$ evolution. For the sake of example, Figure \ref{DCgrow}
shows the $\Delta_{co}$ dependence on $\tau$ for the 4 greatest values
of $z_{\delta=0.01}$ in the previous Figure.

The evolution found depends on $\tilde \tau$, when $\beta_{eff}$
decrease modifies the coefficient in dynamical equations. A visual
confirm comes from Figure 1, where the location of the abrupt
$\beta_{eff}$ decrease is shown.

Before concluding this Section we can also report our estimate on the
relevance of non--coDM components, in the spherical growth analysis.
When assuming them not to be involved in the dynamical process and to
keep homogeneous, the growth rate found has a slight decrease. As
expected, the discrepancy increases towards lower redshifts, but
however keeps within $\sim 3$--$4\, \%$. Let us outline, once more,
that this is an overestimate. Accordingly, results obtained by
considering only coDM fluctuations are however significant, bearing
more than a qualitative validity.

\subsection{Virialization}

As shown in Figure \ref{evosphe}, an {\it ideal} top--hat expands and
eventually recontracts down to a relativistic regime, as indicated by
the black curves.

Since early treatments of top--hat evolution (\cite{press}, hereafter
PS), it was outlined that minimal deviations from sphericity, scarcely
mattering during expansion, become determinant during recontraction,
so leading the system to virialization. In the case of coDM, there is
a specific argument strengthening this expectation, which will be
discussed in Section 6, herebelow.

In order to evaluate when virialization is approached, we however need
to evaluate the potential energy and the kinetic energy $T_{co}(R)$;
we shall do so by treating all components but coDM as homogeneous, so
that the only relevant kinetic energy contribution reads
\begin{equation}
T_{co}(R) = {3 \over 10} M_{co} \left( dR \over dt \right)^{2}~.
\label{Tc}
\end{equation}
Being then $
  { dR /dt} = (1 / a){ dR / d\tau} = \dot c + (\dot a
    / a) \, c\, , $
\begin{equation}
  2 \times {5 \over 3} {T_{co} \over M_{co}} \times { \bar \tau^2
    \over \bar c^2 } = \left(x' + {1 \over u} x \right)^2~,
\label{kinetic}
\end{equation}
with $'$ indicating differentiation in respect to $u$.

The potential energy is then made of two terms: (i) coDM
self-interaction; (ii) coDM interaction with the background of all the
other components. Therefore, in agreement with
\cite{mainini2005,mabo}, we obtain
\begin{align}
\nonumber
\noindent
{U(R) \over M_{co}} &= -{3 \over 5} G {[\langle M_{co} \rangle + \gamma
    \Delta M_{co}] \over R } -{4 \pi \over 5}G \rho_{back} R^2 = \\
  &= -{3 \over 5} \gamma G { \Delta M_{co} \over R } -{4 \pi \over 5}G
  \rho_{cr} R^2~.
\end{align}
and, from here, proceeding as we did for eq. (\ref{eq2}) in the Appendix, 
we finally obtain:

\begin{equation}
  {5 \over 3} {U_{co}(R) \over M_{co}} \times {\tau^2 \over \bar c^2}
  = - h_2\left[{ x^2 \over 2} + {q \over x} (\bar \Delta_{co} -x^3 )
    \right].
\label{potentia}
\end{equation}
Here $h_2 = (8\pi/3) G\rho_{cr} (a\tau)^2$ deviates from unity when
purely radiative expansion is abandoned. In turn, owing to
eq.~(\ref{gstar}), $q = \gamma \Omega_{co}/2 \equiv 1/3 +
1/(4\beta^2)$ exhibits just a mild $\beta$ dependence. The
virialization condition reads then
\begin{equation}
(u x' + {h^{1/2}_2} x)^2 - qh_2(\bar \Delta_{co}/x -x^2 ) -h_2 x^2/2 = 0~.
\label{virial1}
\end{equation}
From the $c_v$ and $\tau_v$ values fulfilling this equation, we then
derive the {\it virial radius} $R_v = c_v a_v$.

\begin{figure}{}
\vskip -2.truecm
\centering
\includegraphics[width=8.9cm]{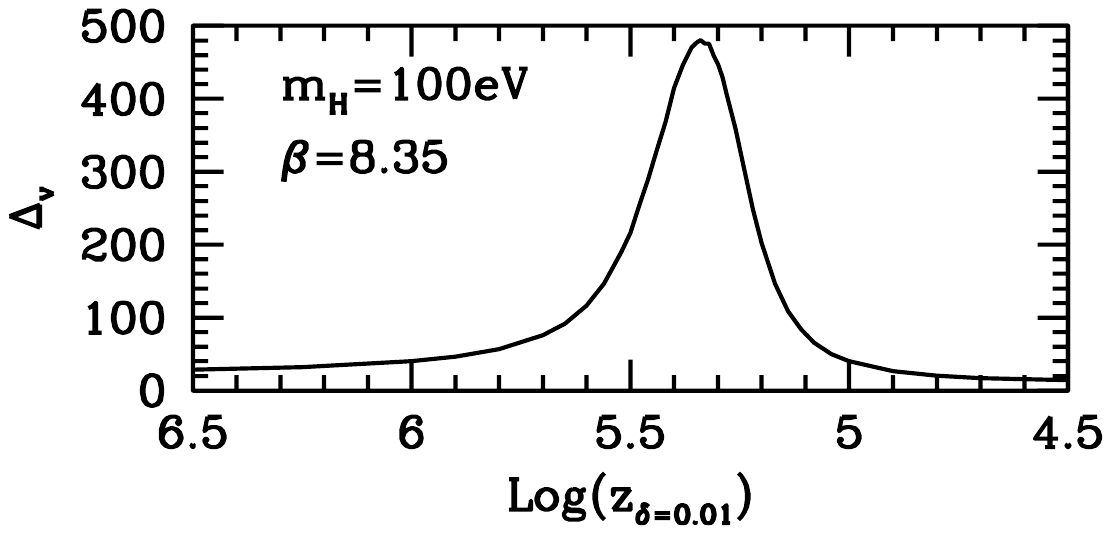}
\vskip -2.5truecm
\vskip -2.9truecm
\vskip -2.9truecm
\includegraphics[width=8.9cm]{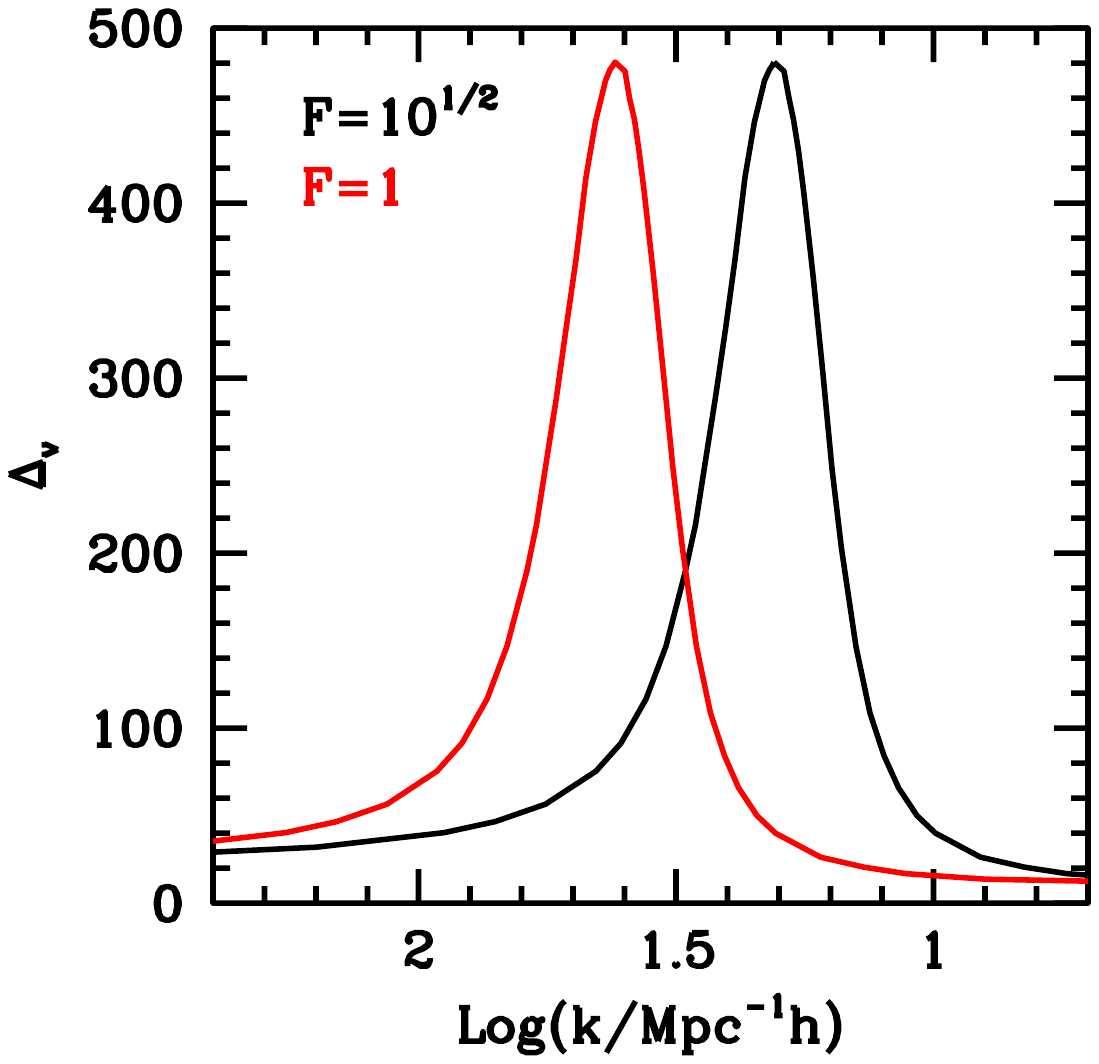}
\vskip -2.7truecm
\caption{Upper plot: The dependence of virial density contrast on the
  redshift when $\delta_{co}=10^{-2}$. Lower plot: Dependence on the scale
  $k$. }
\label{X2art}
\end{figure}

This procedure allows us to work out the virial density contrast
$\Delta_{co,v}$ for coDM fluctuations, as a function of the redshift
when: (i) $\delta_{co}$ has a prescribed value, if, at the horizon,
(ii) the actual fluctuation on that scale exceeded average by a factor
$F$. In Figure \ref{X2art} such dependence is shown for $F=\sqrt{10}$
and 1.
The peak of the density contrast in coDM, $\Delta_{v}$, occurs at
$k \simeq 21.6\, h\, $Mpc$^{-1}$, for $F = \sqrt{10}$~~ ($\simeq 10.4\,
h\, $Mpc$^{-1}$, for $F = 1$).

The related mass values
\begin{equation}
  \label{mp}
 M_P  \simeq {4\pi \over 3} \left( 2\pi \over
k_P \right)^3 \rho_{0,cr} \Omega_{0,co}
\end{equation}
(here $\rho_{0,cr}$ is the present critical density, $\Omega_{0,co}$
is the present coDM density) are $M_P \simeq 3.37 \times 10^7 M_\odot
h^{-1}$ for $F=1$ and $M_P \simeq 2.84 \times 10^8 M_\odot h^{-1}$ for
$F= \sqrt{10}$. According to eq.~(\ref{mp}), here we are taking into
account only the mass $m_H$ for coDM particles. Furthermore, the
infall of other components during the spherical growth is also
neglected. This could imply an error up to $\sim 5\, \%$; the most
significant matter infall, however, is expected to occur at low $z$'s
(see also the Discussion Section).

Let us also outline that average fluctuations entering the horizon are
unlikely to approach a spherical geometry.  Sphericity becomes more
and more likely as $F$ increases. However, fluctuations with greater
$F$ are rarer. Here below the competition between the two effects is
further discussed. Here, let us rather outline that if, e.g., we
take $F=10$, the time taken by the fluctuation to reach an amplitude
$\sim 0.01$ is shorter; accordingly we are referring to a wider
horizon and a greater mass scale. In the early expansion, the mass
scale also exhibits a clear dependence on $\beta$, as $\Omega_{co}
\simeq 1/2\beta^2$.  Accordingly if, e.g., $\beta$ is doubled, the
peak corresponds to a mass smaller by a factor $\simeq 0.25$. This
effect weakens when $\beta_{eff}$ is significantly below $\beta$.  In
the Discussion Section we shall further comment on the reason why the
model(s) used here were selected and how far the parameter choice can
be relaxed.

\section{After virialization}

In previous Sections, the approach to virialization is treated in a
schematic way. As a matter of fact, to settle in virial equilibrium,
the top--hat needs that (tiny) deviations from full homogeneity
existed since the beginning. This requirement, however, is not
different from what \cite{press} claimed to occur, after
recombination, for the evolution of a top--hat fluctuation in baryons
and Dark Matter.

There is however a critical difference between their case and the
present context. \cite{bm2} and \cite{bm3}, in fact, showed that the
vanishing of the virial requires that the average momentum
\begin{equation}
p_v^2 \simeq \gamma G N_{co} m_{eff}^3 (\tau_v)/ R_v~;
\label{pv2}
\end{equation}
here $N_{co}$ is the total number of coupled--DM particles, yielding a
total mass $N_{co}m_{eff}$ within a volume of size $R_v$. Also
\cite{press} require a similar condition, but here $m_{eff}$ exhibits
a time dependence.

In their case, oscillations around virial equilibrium may occur, while
the (conserved) average particle momentum ($p_v$) remains the momentum
yielding virial equilibrium.  On the contrary, here (until
$\beta_{eff}$ is large), $p_v$ soon exceeds the equilibrium momentum
and particles with kinetic energy $p_v^2/2m_{eff}$ are able to
evaporate.

If one assumes a maxwellian distribution, it has been shown that the
fastest particles, while evaporating sooner, are however unable to
produce an average momentum decrease sufficient to recover a temporary
virial equilibrium.  Accordingly, systems virializing with a density
contrast $\sim 28$--30, at high $z$, evaporate within the very {\it
  crossing time}
\begin{equation}
  t_{cross} = 2t_v (qh_2\Delta_v)^{-1/2} \sim 0.7\, t_v~.
  \label{tcross}
\end{equation}
Here $t_v$ is the time when virialization is achieved.  This result
however opens another question: If the (conformal) time when the
virialization condition is attained is $\tau_v$, when does the growth
really stop? If this very stop is expected to take place when the
(conformal) time has grown enough to allow a full crossing to occur,
i.e. at a time $\sim 1.3\, \tau_v$, we shift from $\tau/\tilde \tau
\simeq 24$ to $\tau/\tilde \tau \simeq 31.3 $. As full collapse, if
sphericity is not violated, is expected when $\tau/\tilde \tau \simeq
28 $, does this mean that any spherical fluctuation is doomed to turn
into a BH~? Small violations from sphericity could however prevent
this to occur and, in the next Sections, we actually discuss on the
{\it a}--sphericity needed to stop the collapse.  However, if we
suppose that a nearly spherical fluctuation indeed virializes at a
time $t_v$, we just discovered that, afterwards, the escape momentum
decreases too rapidly. Henceforth, at the end of the growth, {\it
  either a BH is formed or no trapping effect is possible: particle
  simply flow out from the overdensity within a time $\sim 0.7\,
  t_v$.}

As soon as $\beta_{eff}$ weakens, however, the balance between the two
options is expected to change, as the main mass component is the mass
$\tilde \mu$ acquired at the higgs scale, so that coDM particle mass
decrease has a stop.  For spherical systems reaching the virialization
condition with density contrast $\sim 450$--480, close to the maximum,
therefore, the dissolution option seems excluded. They surely may
loose a part of their mass, but the likelihood that contraction
continues towards BH formation becomes huger, also because the very
ratio between virialization and full collapse (conformal) times
decreases from $\sim 1.17$ to $\sim 1.07$--1.08~.

\section{Stability of sphericity assumption}

The critical issue, therefore, seems to be sphericity and its
stability during the contraction stages. The PS approach, when
ordinary cold DM and baryons are involved, clearly forgets all
hydrodynamical effects, which could be sufficient to modify purely
gravitational predictions during the contraction stages. In
particular, they would act on any substructure, even if the physical
fluctuation is spherical.  Hydrodynamics is not the only reason why a
PS collapse is unstable, but its action is surely unavoidable.

A first significant difference, in the coDM case, is that baryons, if
involved, keep far behind coDM evolution; when coDM fluctuations
overcome their top expansion, baryons or other components are still
timidly hinting a linear growth. As we saw in detail, forgetting any
other component besides coDM is then a fair approximation.

There is however a peculiar feature of coDM dynamics, which could
decisively contribute to sphericity disruption. To this end, let us
compare the two components of the force slowing down the growth of a
density excess, in respect to cosmic expansion, and then causing
recontraction. According to eq.~(\ref{eq1}), in Appendix A, the
gravitational force reads
\begin{equation}
  F_G = -G\frac{\Delta {\cal M}_{co}}{ac^2}~,
\end{equation}
with a suitable expression for $\Delta {\cal M}_{co}$, taking into
account the boosted gravity; this force is directed towards the
gravity center.  In top of that, we have an {\it extra push}
\begin{equation}
  F_P = C\dot\Phi \dot c ~,
\end{equation}
directed as particle velocities. In average, for a spherical density
enhancement, both forces are indeed radially directed.

We can however use the solutions of dynamical equations, to compare
them, once normalized as in eq.~(\ref{eq3}) of Appendix A. This is
done in Figs.~\ref{gravpush9} or \ref{gravpush5} if the redshift when
$\delta_{co}=10^{-2}$ is $10^9$ or $10^{5.35}$, i.e. for fluctuation
scales reached by the horizon either when CI expansion is still going
on or during the onset of the $\beta_{eff}$ decrease due to higgs
screening.
\begin{figure}{}
\vskip -2.truecm
\centering
\includegraphics[width=8.5cm]{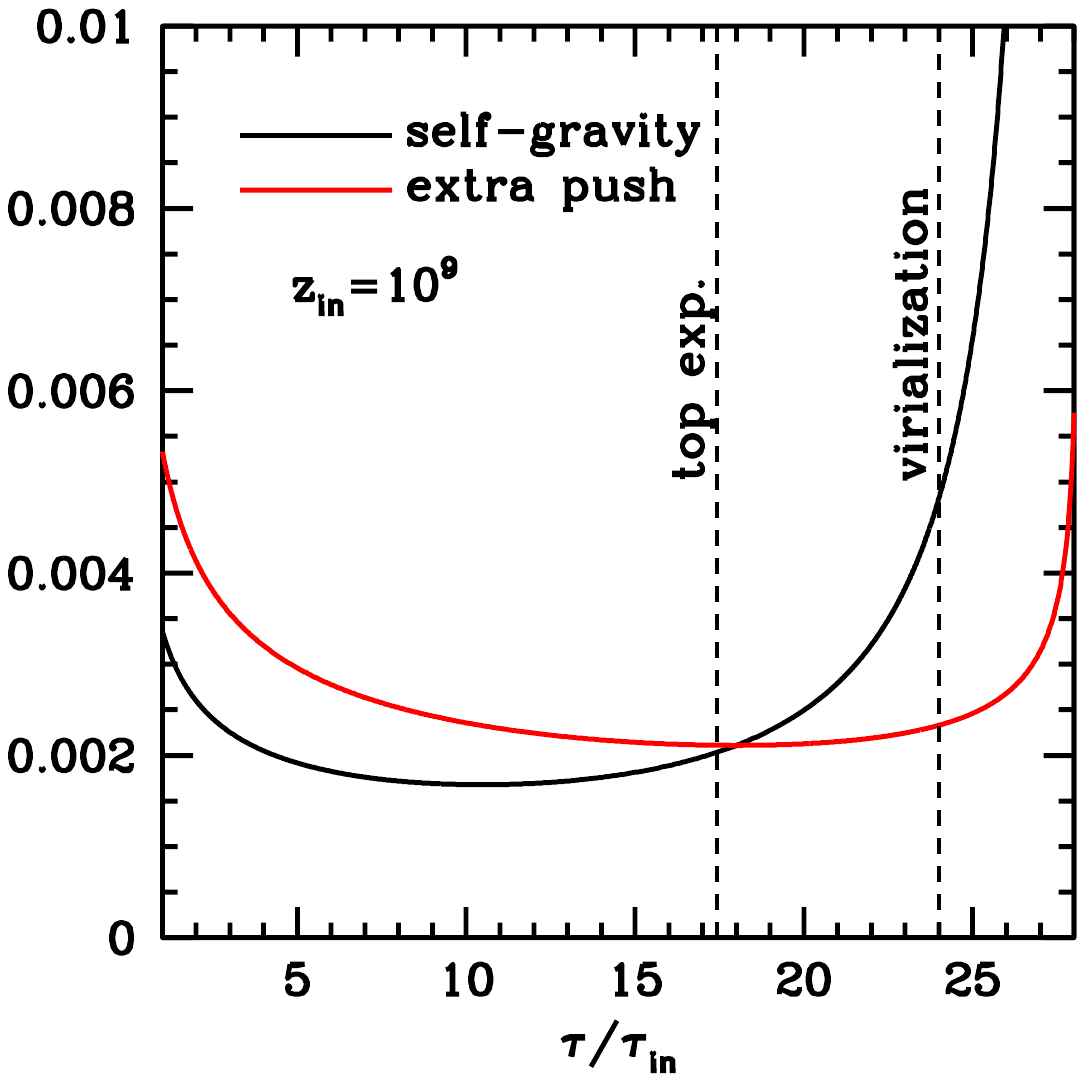}
\vskip -2.4truecm
\caption{Boosted gravity and extra push, for top--hat fluctuations
  starting with $\delta_{co}=10^{-2}$ at $z=10^9$, when still in the
  high--$z$ regime.}
\label{gravpush9}
\end{figure}   
\begin{figure}{}
\vskip -2.truecm
\centering
\includegraphics[width=8.5cm]{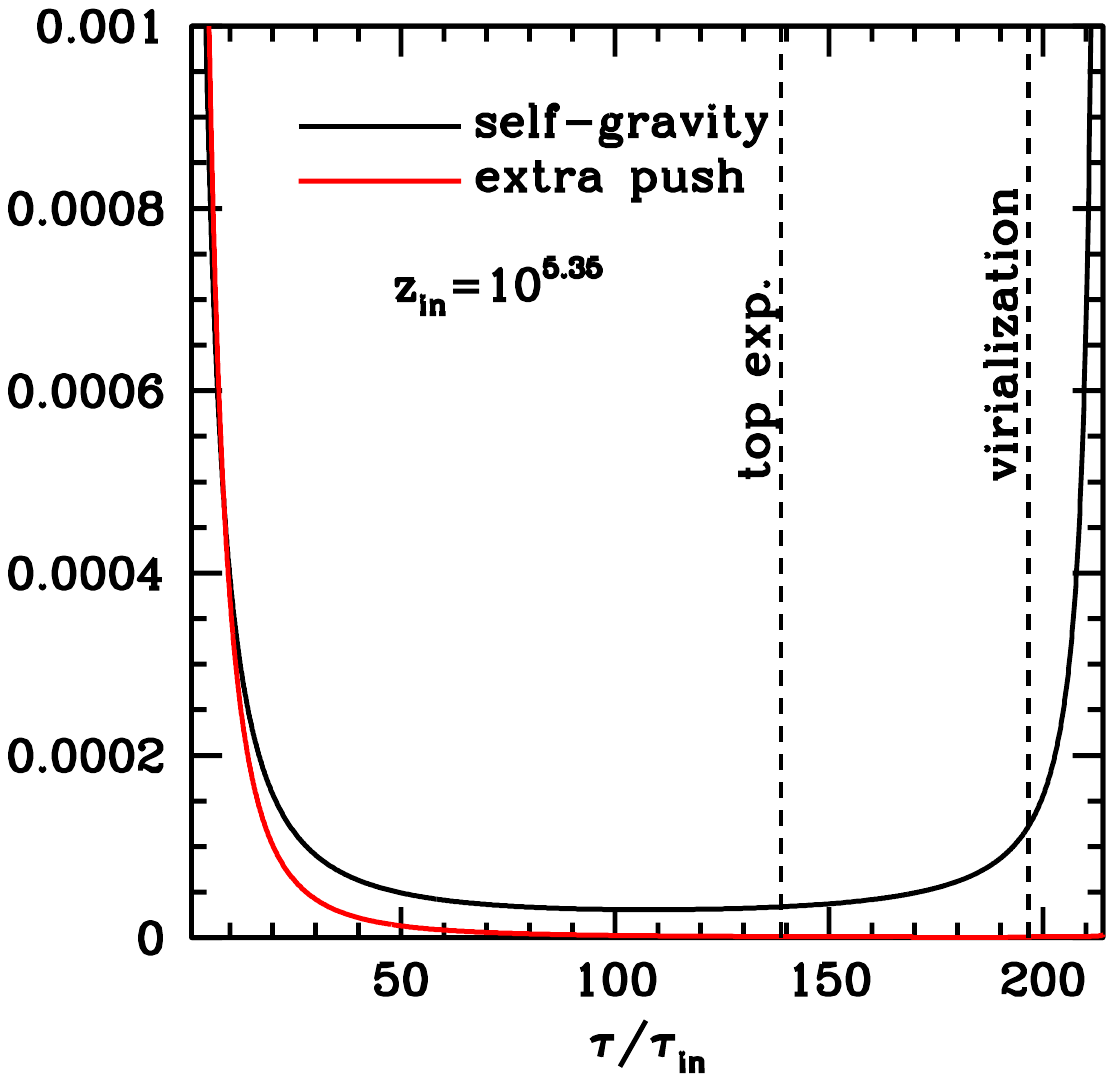}
\vskip -6.8truecm
\includegraphics[width=8.5cm]{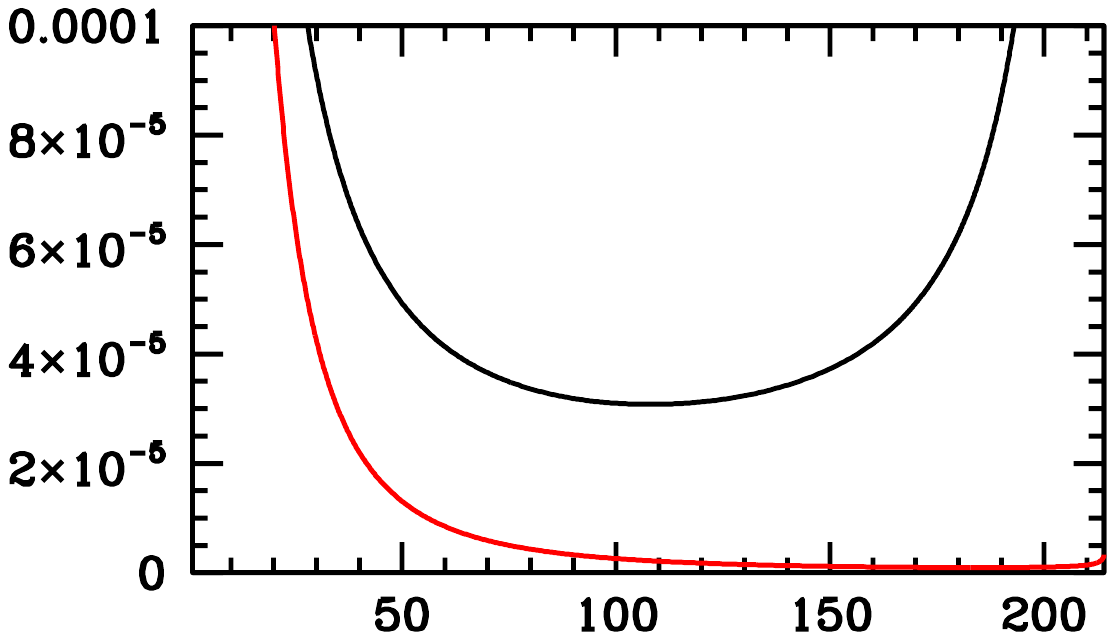}
\vskip -2.7truecm
\caption{Boosted gravity and extra push, for top--hat fluctuations
  starting with $\delta_{co}=10^{-2}$ at $z=10^{5.35}$, i.e., in the
  proximity of the peak in Fig.~8. The lower 10$\, \%$ of the upper
  plot is magnified in the lower plot.}
\label{gravpush5}
\end{figure}   

These plots start at the time when $\delta_{co}=10^{-2}$ and end when
recontraction approaches the speed of light, so indicating that the
non--relativistic regime is over. Quite in general, they make clear
that extra push intensity can approach self--gravity. As is expected,
Figures \ref{gravpush9} and \ref{gravpush5} describe fairly different
evolutions, also because the very top virial density contrasts are
different.

In fact, all through Figure \ref{gravpush9}, $\beta_{eff}$ keeps close
to $\beta$ and the mass acquired by the $\psi$ field at the higgs
scale keeps negligible.  Enhanced gravity and extra push then keep
similar intensities all through the process.

If (possible) sphericity violations are mild, the two forces, directed
towards the gravity center and in the velocity direction, act
coherently. Possible {\it a--}sphericities, however, would make these
forces discrepant, and this is a fair reason, intrinsic to coDM
nature, bursting deviations from sphericity, even though initially
small. If we suppose that coDM feels no other effective force apart
these ones, we then face a precise quantitative problem: how large
deviations from sphericity are ``tolerable''?

Figures \ref{gravpush5} then describe a growth leading to a top
density contrast $\sim 500$. Here, the extra push is significant only
at the very beginning; then, it even exceeds self--gravity. Both
forces then fade, after yielding a strong initial headway, as
$\beta_{eff}$ becomes significantly smaller than $\beta$: gravity,
then, is no longer enhanced as the $\psi$ field approaches a constant
mass. Clearly, the transition from variable to constant mass occurs
while expansion is running. It is however clear that, if sphericity is
not disrupted during the initial stages, the mismatch between gravity
and extra push directions should not be a possible cause for a later
mixing up, leading to virialization.

Assuming that deviations from sphericity are similarly distributed,
for the two scales considered, it seems then clear that, in the latter
case, contraction towards BH formation is favored. In Appendix B, we
debate deviations from sphericity as a function of $F$.

\begin{figure}{}
\vskip -2.truecm
\centering
\includegraphics[width=8.5cm]{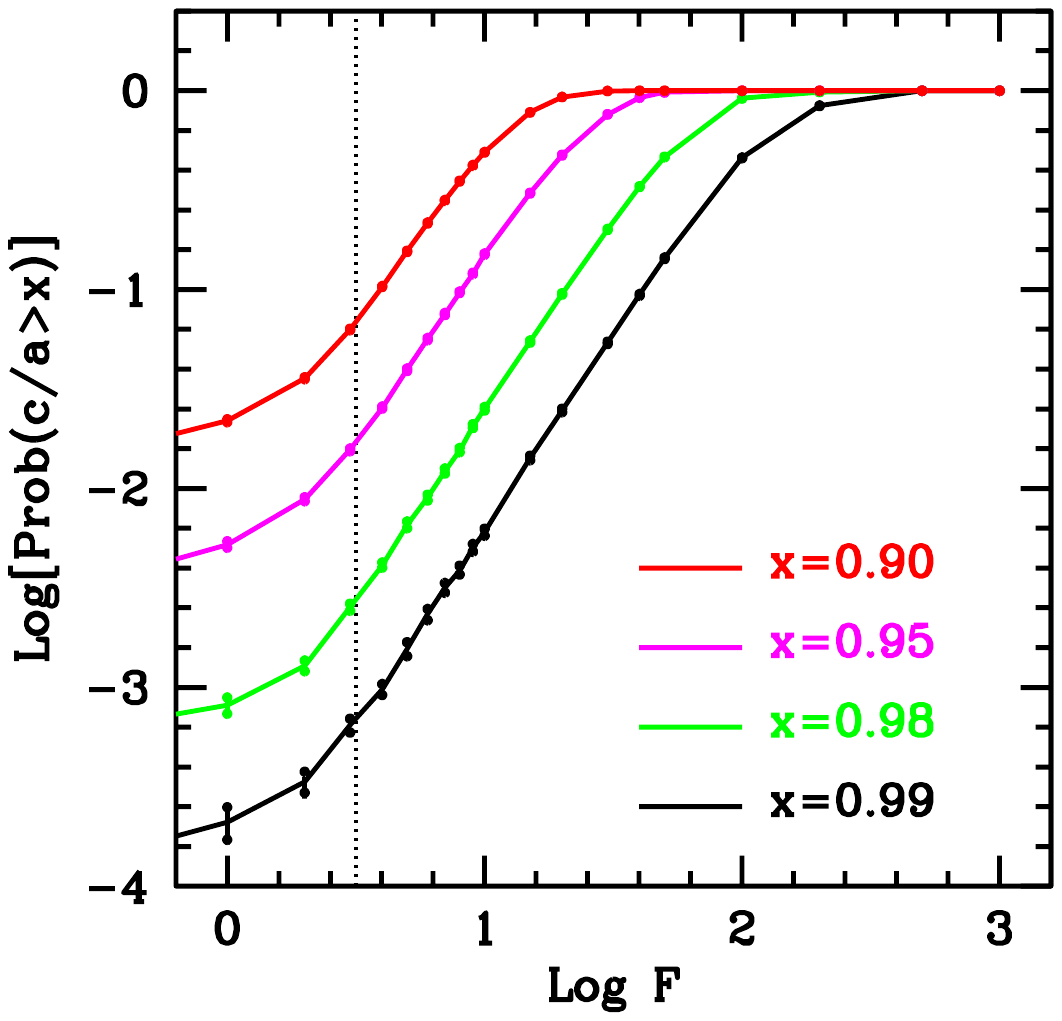}
\vskip -2.4truecm
\caption{Probability for the ratio $c/a$ (maximum/minimum ellipsoidal
  radius) to keep below 4 possible fixed limits, for increasing $F$
  values. Each estimate includes a 2--$\sigma$ errorbar. The vertical
  dotted line is for $F= \sqrt{10}$.}
\label{marinplot}
\end{figure}   
As expected, the fraction of fluctuations with a quasi--spherical
shape increases with $F$ (see Figure \ref{marinplot}). The likelihood
of fluctuations with assigned $F$ is however expected to decrease
$\propto \exp(-F^2/2)$. For instance, fluctuations with $F=\sqrt{10}$
are $e^{-5} \simeq 10^{-2.2}$ times less frequent than fluctuations
with $F=1$. In turn, the fraction of fluctuations with $c/a > 0.99$
($a > b > c$~ are the 3 major axis of the fluctuation) increases by a
factor $\sim 10^{0.7}$. The (somehow unexpected) result of this
comparison is that ``spherical'' fluctuations with m.s.a. amplitude
are much more frequent than fluctuations whose amplitude exceeds
average.

It should be also outlined that the requirement that $c/a > 0.99$ is
only a necessary condition for the fluctuation to approach a spherical
geometry. See Appendix B for further details on this point.

\section{Pancakes}
Fluctuations approaching a spherical geometry are however an
exception. Let us then approach the question of how typical
fluctuations in coDM evolve, by considering the coDM density
distribution $\rho_{co}(\tau,{\bf r})=m(\tau) \times n(\tau,{\bf r})$
in the neighborhood of a given point ${\bf x} = a{\bf r}$, whereabout
an overdensity is located. Once the principal ellipsoidal axes are
determined, we can face the problem in a Zel'dovich approximation, by
requiring that
\begin{equation}
 n({\bf r}) \, a^3\, \prod_i[1+\alpha_i {\cal G}(a)] = {\rm const.}~;
\end{equation}
here $\alpha_i$ are negative coefficients, the product being extended
over the 3 ellipsoidal axes. The linear algorithm tells us that ${\cal
  G} \propto a^{\alpha}$ and we shall keep to the case $\alpha=1.6$;
being $\delta_{co} \simeq {\cal G}\sum_i \alpha_i \propto a^{1.6}$, we
soon obtain a fair translation from the eulerian to the lagrangian
picture.
  
The turnaround time, when a coefficient $[a+\alpha_i a^{2.6}]$ shifts
from increase to decrease, is obtainable by requiring
\begin{equation}
  {d \over d\tau} [a+\alpha_i a^{2.6}] = \dot a \times
  [1+2.6\, \alpha_i a^{1.6}] = 0~,
\end{equation}
so obtaining
\begin{equation}
  \alpha_i a^{1.6} =  -0.385~~~~~{\rm i.e.}
  ~~~~~~ a = \left[ 1 \over 2.6(-\alpha_i) \right]^{0.625}
\end{equation}
and
\begin{equation}
a+\alpha_i a^{2.6} = a \times 0.615~;
\end{equation}
accordingly, for a density contrast $\Delta_{t.a.} \simeq 1/0.615 =
1.6$ and a fluctuation amplitude
\begin{equation}
  \delta_{co}(\tau_{t.a.}) \simeq 0.6~,
\end{equation}
still almost linear, turnaround occurs. Non--linearity corrections,
therefore, can be expected to be small.

In a standard Zel'dovich approach, the successive compression is
expected to burst pressure, so causing fragmentation. On the contrary,
coDM knots are pressureless and their particles are (almost)
noninteracting.  Rather, we expect a key role to be played by the
mismatch between enhanced gravity and extra push directions, acting
towards the ellipsoid center and, in average, along the contraction
axis, respectively.  Accordingly, kinetic energy is preserved but
particle velocities are randomly diverted so that a local virial
equilibrium can be easily approached, and the growth stops. After a
short while, then, as in the spherical case, evaporation starts and
the whole fluctuation energy is dissipated into heat.

Notice that previous evaluations referred to $ n({\bf r})$ rather than
$ \rho_{co}({\bf r})$. For CDM or baryons the two options would be
equivalent. For coDM, on the contrary, until the particle mass
$m_{eff}(\tau) \propto \tau^{-1}$, we must refer to (conserved) number
densities.

One dimensional virial equilibrium however requires that the average
particle momentum is
\begin{equation}
  \label{vir}
  \langle p^2 \rangle \simeq \gamma G N_{co} m_{eff}^3/R~.
\end{equation}
Here we can take $R \sim (a-|\alpha_i| a^{2.6})2\pi/k$ to be the
thickness attained by the growing fluctuation, $N_{co}$ is the number
of coDM particles involved in the growth.

In the spherical case, the equilibrium set by eq.~(\ref{pv2}) is
reached when $a \simeq 1.4\, a_{t.a.}$ and $R$ has decreased by a
factor $\simeq 0.52\, $ to reach a density contrast $\sim
27$--28. (For the sake of comparison, in the PS case the scale factor
increased up to $a \simeq 4\, a_{t.a.}$ and, namely, the virial
density contrast is $\Delta \sim 180$.) The density fluctuations of
coDM particles, therefore, undergo both a fast growth and a rapid
virialization, followed by dissolution.

In the one--dimensional collapse, while the $\delta_{co}$ growth is
reasonably approximated by linear estimates, until turning around, the
expected law $\delta_{co}(\tau > \tau_{t.a.} )$ is admittedly hard to
predict.

\section{Transfer functions}

Accordingly, it is uneasy to predict the effects of coDM
\begin{figure}{}
\vskip -2.1truecm
\centering
\includegraphics[width=8.1cm]{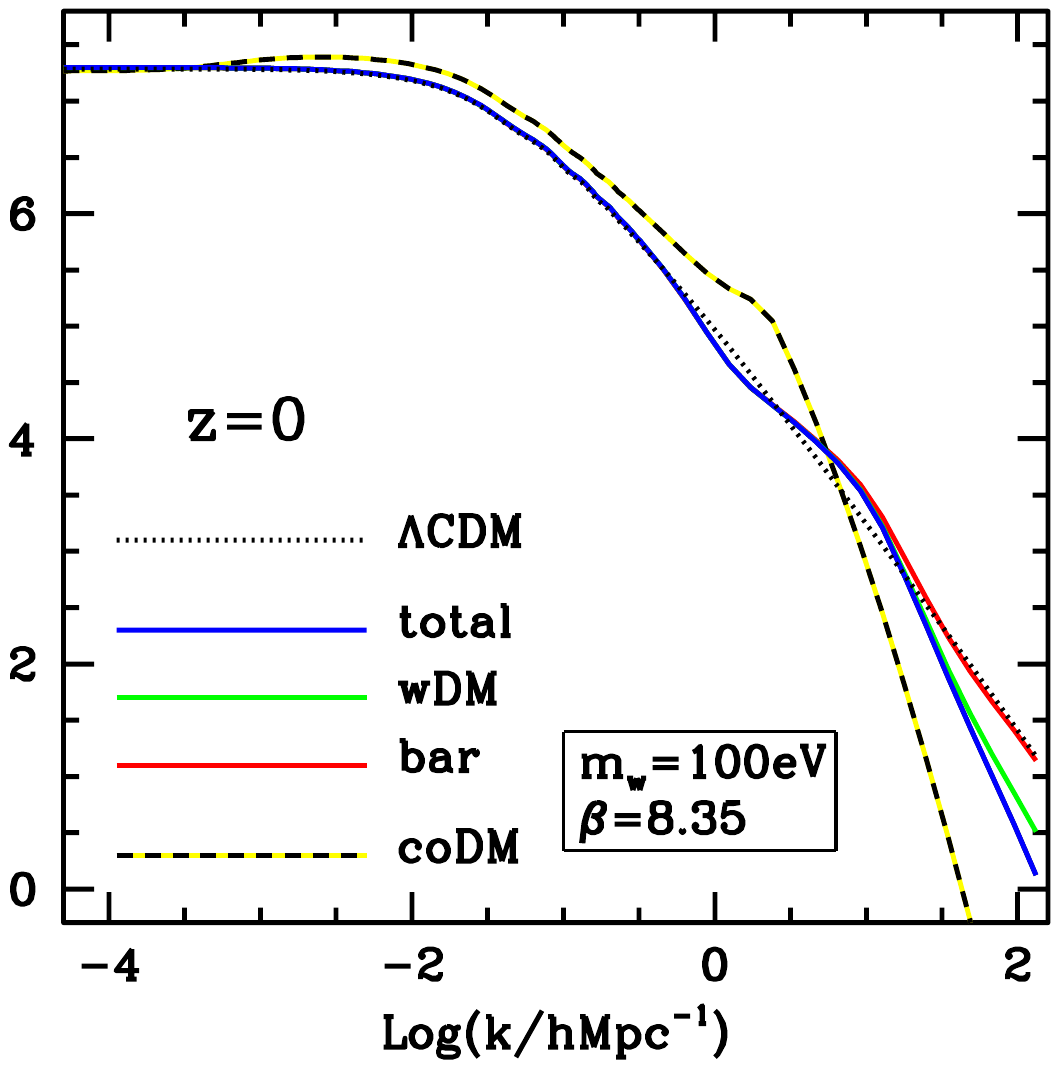}
\vskip -2.1truecm
\caption{Transfer functions of SCDEW (parameters in the box) and \LCDM
  compared at $z=0$.}
\vskip -.3truecm
\label{tfplot}
\end{figure}   
non--linearities on other components. This is a critical issue, for
coDM non--linearities leave an imprint on baryons and wDM
distributions. As a tentative option, we model the effects of
$\delta_{co}$ on other components, by letting it evolve according to
the linear algorithm until $\delta_{co} \simeq 3$, at a time $\tau_3$,
and assuming a later exponential decay, due to post--virial
dissolution; e.g.:
\begin{equation}
  \label{added}
  \delta_{co}(\tau > \tau_3) = \delta_{co}(\tau_3)
  \exp[-\alpha(\tau/\tau_3-1)^n]
\end{equation}
(here we selected $\alpha=1.5$; $n$ is then derived from continuity
requirements at $\tau_3$). Here, a virial density contrast
$\Delta_{1,co} \simeq 4$ (2.5 times the turn--around density contrast)
is assumed, which could be an
\begin{figure}{}
\vskip -2.1truecm
\centering
\includegraphics[width=8.1cm]{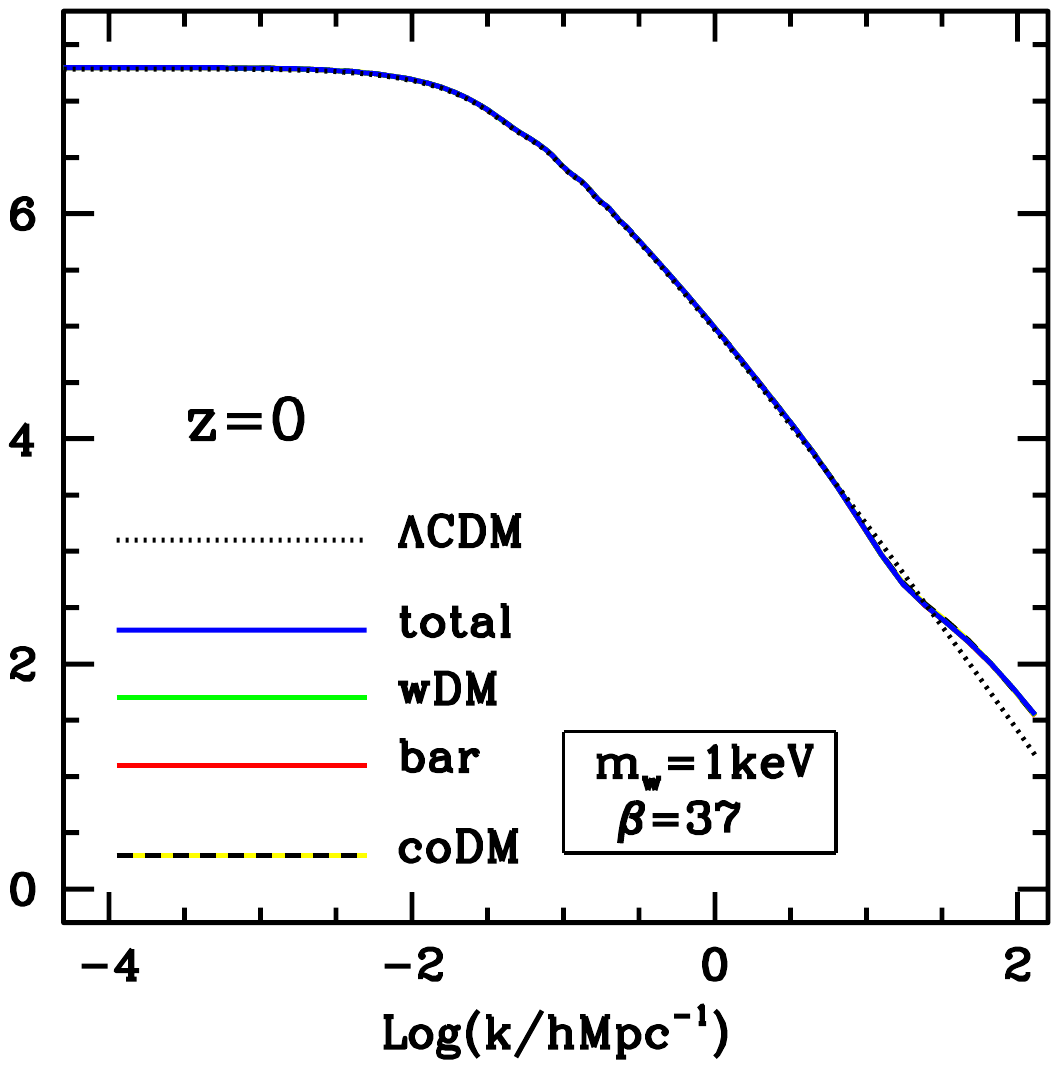}
\vskip -4.truecm
\includegraphics[width=8.1cm]{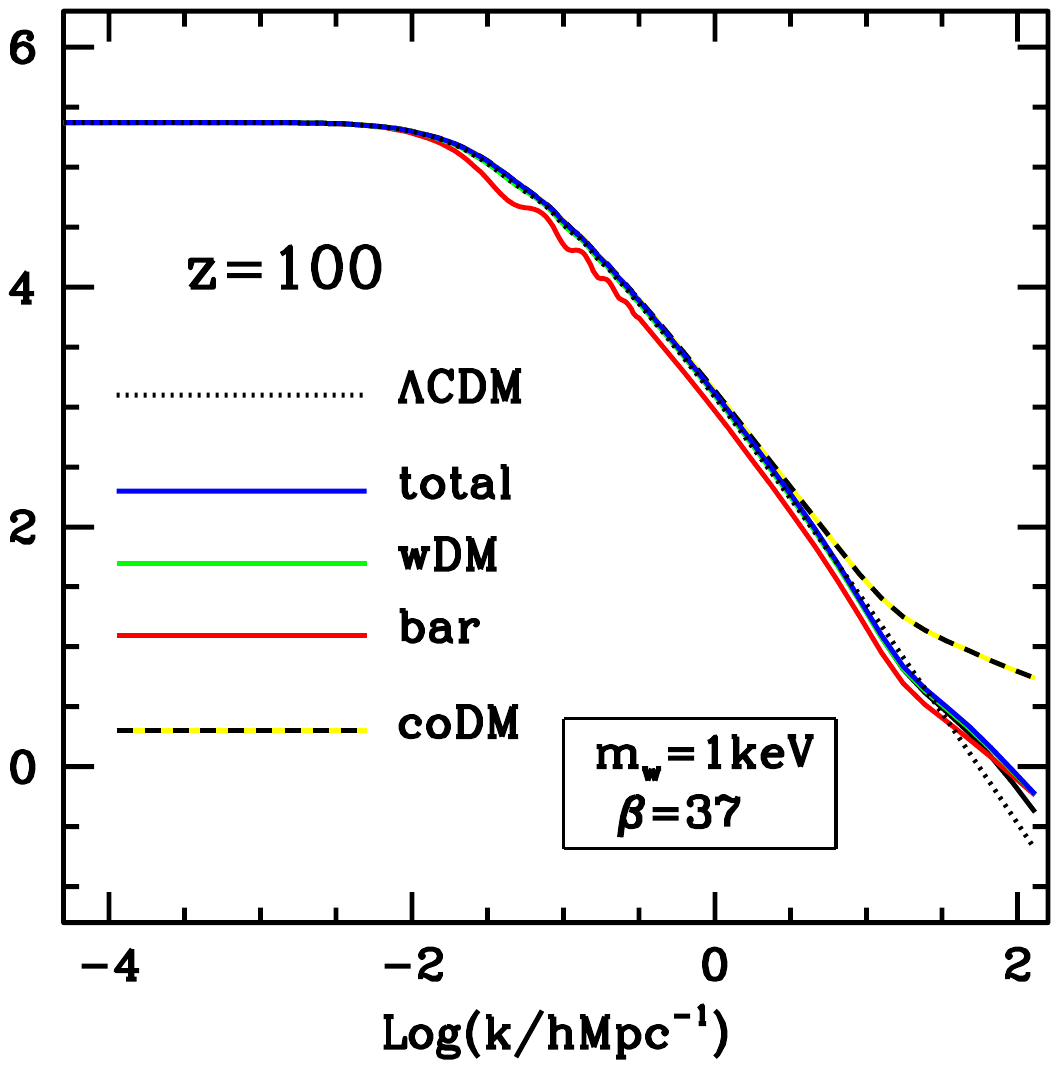}
\vskip -2.1truecm
\caption{Transfer functions of SCDEW (with $m_H=1\, $keV) and \LCDM
  compared, at $z=0$ (when they are quite close) and 100 (where they
  still exhibit some differences).}
\vskip -.3truecm
\label{tf0100plot}
\end{figure}   
underestimate; that the time needed to start dissolution is the time
needed to reach $\delta_{co} = 3$, on the contrary, could be an
overestimate. Until more precise results are obtained through ad--hoc
simulations, the best we can do, probably, is assuming a mutual
compensation of these approximations.

A modified {\sl cmbfast} program, allowing us to predict transfer
functions and spectra, when making these (or close) assumptions, has
also been produced. A discussion of the (slight) spectral dependence
on parameters (one of them is $\alpha$, here set to 1.5) enclosed --or
to be added-- in the expression (\ref{added}), will be provided in a
forthcoming paper.

For the sake of example, in Figure \ref{tfplot} we compare the
transfer function of a SCDEW model, with $\beta=8.35$ and $m_w=100\,
$eV, with \LCDM, at $z=0$. Model parameters ($\Omega$'s, $h$, $n_s$,
etc.) are the same, for the two cosmologies, coinciding with those
assumed in the NIHAO simulation set (e.g., \cite{nihao1, nihao4,
  nihao5}). The coDM spectrum exhibits significant deviations from
\LCDM. But one should remind that it mostly accounts for $1/2\beta^2
\simeq 0.7\, \%$ of the critical density. Its contribution is however
taken into account in the total spectrum (blue curve), exhibiting
deviations $< 10\, \%$ up to $k \simeq 30\, h\, $Mpc$^{-1}$, i.e. up
to $\sim 10^{10}M_\odot h^{-1}$, well inside the non--linear spectrum;
high--$z$ data fitting could be a more serious test for this
model. Most computations in this paper were however based on this
model and, in the discussion Section, we recall why it was selected.

In general, however, SCDEW spectra are increasingly closer to \LCDM
when greater $\beta$ and $m_w$ are selected; e.g., in the Figures
\ref{tf0100plot}, the model with $\beta=37$ and $m_w=1\, $keV is
considered. At $z=0$, this model almost overlaps \LCDM, with a tiny
lack of power at $k \simeq 10\, h\, $Mpc$^{-1}$ and some power in
excess for $k > 50\, h\, $Mpc$^{-1}$; shifts are however within
4--5$\, \%$. At $z = 100$, discrepancies are slightly more
significant. The coDM component, in particular, has still some extra
power at large $k$'s, while baryon oscillations are visible on the
baryon component only.

Let us recall that wDM particle masses, in these models, are 100 or
1000 eV. With such masses, a WDM model transfer function exhibits a
cut around $k=1$ or $10\, h\, $Mpc$^{-1}$, i.e. $\sim 10^{14}$ or
$10^{11} M_\odot h^{-2}$, respectively.  The role of coDM in rising
high--$k$ spectral components is therefore essential.

This takes us back to the problem of improving the approximation in
eq.~(\ref{added}). Let us however outline that only wide changes in it
would substantially modify the similitude between \LCDM and SCDEW at
high $k$. Furthermore, the basic issue is that, when coDM effects on
baryons and wDM of coDM are evaluated, two possible traps are to be
avoided: (i) Using a standard linear program, letting $\delta_{co}$
grow forever, as thought linear equations held also when
$|\delta_{co}|$ approaches and exceeds unity.  (ii) Testing the
effects on baryons and wDM due to a spherical top--hat (or similar)
coDM fluctuation growth (and dissolution).  Figure \ref{marinplot}
shows in detail how rare are spherical fluctuations, so that the story
of their evolution is unrelated to average fluctuation evolution. An
important point, when trying to improve the expression (\ref{added}),
is also accounting for the contribution of negative fluctuations.

Let us however finally outline that no similar problem exists when
transfer functions are evaluated in most different cosmologies. Early
nonlinearities are a specific feature of SCDEW. In turn, it is thanks
to them that SCDEW can open new perspectives for early BH formation.

\section{Discussion}

SCDEW models have the ambition to challenge \LCDM as concordance
cosmology. More precisely: they aim to show how \LCDM features can be
recovered, by avoiding fine tuning and coincidence problems.

If \LCDM is ``just'' an excellent effective model, we expect that the
underlying physics will carry along new parameters. SCDEW involving
extra parameters, therefore, is hardly a point against it. The true
points is whether, by fixing them, we recover \LCDM or even go beyond
it, not only avoiding logical conundrums but also fixing some open
quantitative questions.

The $\beta$--coupling of SCDEW models could be seen as a remnant
  of the interaction between a primeval inflatonic field and other
  cosmic component. When the higgs scale is attained, however, this
  residual force
almost vanishes, without invoking any {\it ad--hoc}
mechanism. Moreover, one does not need to assume a specific $V(\Phi)$
expression to account for $\Phi$ self--interaction; it is sufficient
to indicate the redshift $z_{kp}$ when $V(\Phi)$ starts to exceed
$\Phi$ kinetic energy density.  Accordingly, $z_{kp}$ tuning just
replaces the tuning of the DE density parameter $\Omega_d$.

More severe constraints on $V$ expressions might derive from its role
in inflation.  Mutual constraints between inflationary dynamics and
$\Omega_d$ could arise from that. This is clearly still an open point.

In a SCDEW cosmology, all cosmic components keep significant densities
all through cosmic expansion, including inflation and
today. Admittedly, there is an exception: baryon density and its being
comparable with other densities; an open question shared with any
other cosmological scenario.

These arguments could be made in support of SCDEW cosmologies even in
earlier analysis. Two new points were added here, both related to
early coDM non--linearities: (i) How to treat their effects on other
cosmic components, so to approach a prediction on high--$k$ SCDEW
spectra. (ii) A possible role of SCDEW dynamics in producing large
scale Black Holes.

\subsection{SCDEW high--$k$ spectra}
The high--$k$ range corresponds to mass scales below the average
galaxy mass scale. Being highly non linear today, the main tool to
constrain predictions are high--$z$ observations.

First of all, here we outline that SCDEW rises a technical
``problem'', that no other cosmology faces.  Then, we provide an
approximate analytical solution. Admittedly, it is not yet fully
satisfactory. Two specific ``traps'' are however to be avoided: (a) A
standard linear program, in this $k$--range, lets $\delta_{co}$ grow
forever, even when $|\delta_{co}|$ approaches and exceeds unity:
extending linear results on such $k$--range is surely wrong and, at
first sight, might seem an under--evaluation of its effects. (b)
Testing coDM effects on baryons and wDM by using a spherical
fluctuation growth (and dissolution) is also wrong: for average
amplitude fluctuations, spherical fluctuations are a black swan.

The approximated expression suggested here, which can be better fixed
by future analysis, is a conceptual step forward in respect to the
options (a) or (b). In particular, it shows that an extension of
``linear'' program results is not a sort of under--evaluation, it is
just plainly wrong.

\subsection{Early BH formation}
As far as the (ii) point is concerned, let us recall that recent
observations at $z > 6$ (see, e.g.,
\cite{fan,willot,mortlock,venemans,wu,banados1,banados2,banados3,matsuoka,tang,chehade})
do reveal that SMBH's (super--massive BH) already existed when the
cosmic age was $\sim 10^9$y. Their masses approach or exceed $10^9
M_\odot$ and various attempts have been made to justify their
existence. In the standard approach (see, e.g.,
\cite{volonteri,volonteribello,haiman,latifferrara})
SMBH are tentatively explained, by assuming the existence of BH seeds,
with mass $\sim 400\, M_\odot$, since $z \sim 15~.$ They should then
coalesce or be subject to accretion. More in detail, the most popular
options are: (i) The so--called DCBH (Direct Collapse Black Hole),
i.e., the collapse of a protogalactic gas cloud, metal free. (ii) The
core collapse of ultra--massive stars. (iii) The collapse of dense
nuclear star clusters.

None of these options is devoid of problems, as well as the very
formation of their seeds. Much work is in progress in this field and
it is even possible that each one of above three options is viable, in
suitably different contexts.

Naively, the direct formation of BH's over scales $M_{BH} \sim 10^8
M_\odot$ could be the best product a {\it deus ex machina} could
provide. The scale $M_{BH}$ is however set by the assumption of a wDM
particle mass $\sim 100\, $eV, therefore neglecting the primeval time
dependent mass scale $\mu \exp(C\Phi) \propto \tau^{-1}$. At the time
when PBH might have formed, however, the latter mass component was
still non--negligible. When enclosed in a BH, coDM particle masses
might still evolve in time; but time coordinates inside and outside
BH's are different; altogether, it is unclear how the observed BH mass
could evolve in time. This point does not modify the expected o.o.m.
of BH masses, but could change their values up to some 10$\, \%$'s.
Another reason why BH mass could be greater is the accretion of
baryons and wDM during the late collapse stages. This option was
however enclosed in the computation of this paper, and causes a
correction never exceeding a few percents.

More severe problems could be caused by accretion in the epoch between
recombination and reionization due to Pop III stars. Not so much
because of the amount of matter then accreted, but because of the
radiation emitted, which risks to cause an early reionization.
Clearly, this point requires a more detailed analysis, that we
postpone to further work.

Let us however keep to the mass at BH formation, by taking into
account just $m_H$ masses. We then considered a sequence of SCDEW
models, starting from $\beta=8.35$ and $m_H=100\, $eV, and
simultaneously increasing $\beta$ and $m_H$, so to keep close values
for $\rho_w$ and $\rho_{co}$, during the early CI expansion.
For $m_H=100\, $eV, it is $\rho_w/\rho_{co} \simeq 0.9$.  Another
model we consider in some detail is $m_H=1\, $keV--$\beta=37$,
yielding a similar primeval density ratio.

This assumption is in agreement with our conjecture that wDM and coDM
own a strictly related origin being, somehow, the $\beta$--charged
and neutral states of the same particle. Of course, instead of a ratio
0.9, we could fix any value close to unity. Results are however just
marginally modified by any such variation.

\begin{figure}{}
\vskip -2.1truecm
\centering
\includegraphics[width=8.1cm]{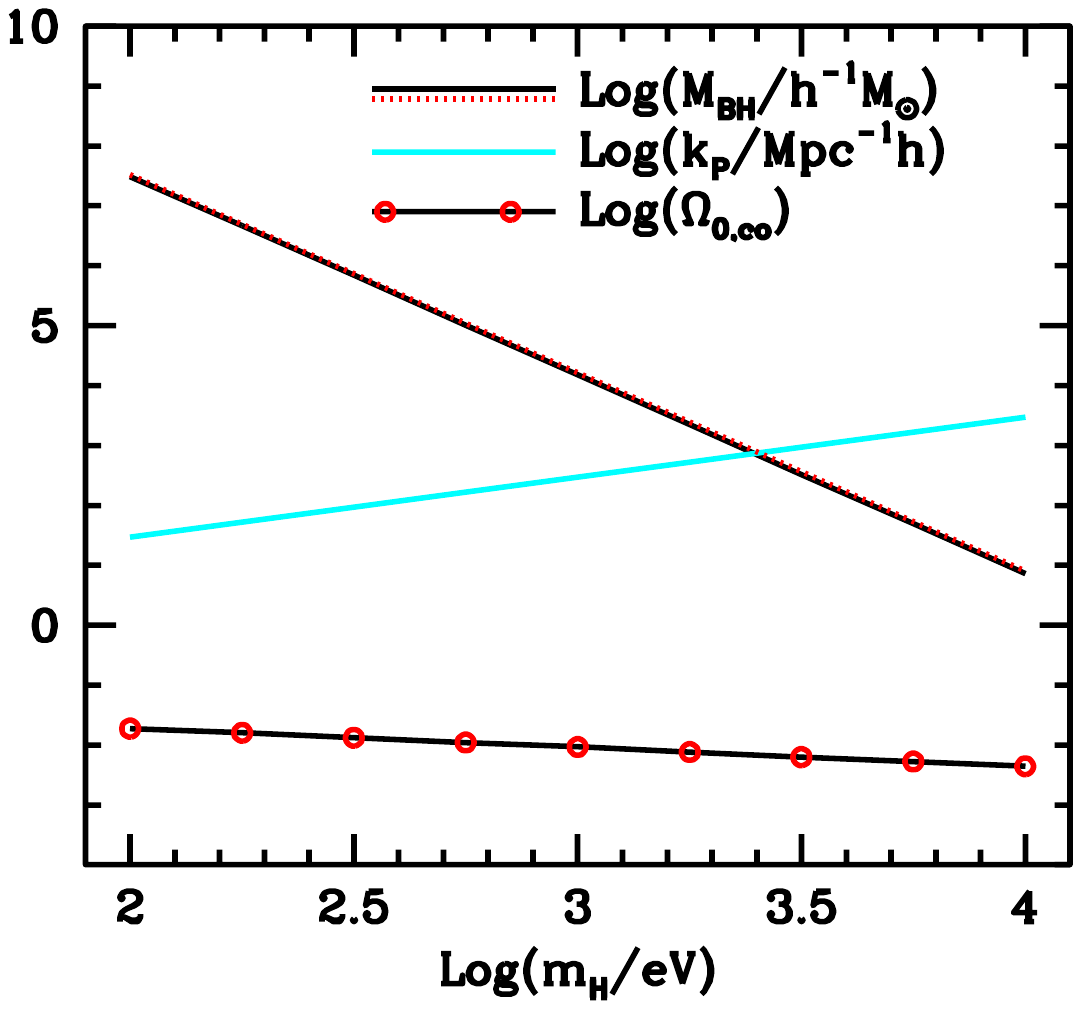}
\vskip -2.5truecm
\caption{Dependence of PBH masses on asymptotic mass value.  The plot
  is obtained by taking values of $\beta$ yielding close densities of
  wDM and coDM at high $z$ (see text).
  In this way the upper black curve is obtained, which can be
  approximated by the red straight line (eq.~1).
}
\vskip -.3truecm
\label{rho0co}
\end{figure}

We then try to apply eq.~(\ref{mbh}) to find the dependence of
$M_P$ (and, therefore, $M_{BH}$) on $m_H$, by taking into account
the corresponding values of $\Omega_{0,co}$. Such values are obtained
from the background program, whose results are plotted in Figure
\ref{rho0co} (black curve with round circles).
In the same Figure we also plot the $m_H$ dependence of $k_P$ (cyan
curve), as well as the resulting dependence on $m_H$ of $M_{BH}$
(black curve). Overimposed onto it a dotted red line yields the
expression (1).

Let us now outline that the choice of a model with $m_w \sim 100\, $eV
was not casual.  In previous work, it allowed us to ease problems that
N--body simulations of \LCDM exhibited, typically below the galactic
mass scale: the abundance of MW and M31 satellites, the flattening of
dwarf profiles, and also, possibly, the distribution of
concentrations.  N--body simulations run by \cite{maccio2015}
confirmed such expected result.

Hydro simulations of \LCDM, as those run for the NIHAO program (see,
e.g., \cite{nihao1,nihao4,nihao5}), showed that these very problems
can find a solution within the \LCDM paradigm. This however requires a
tuning of specific baryon physics parameters, while the concentration
distribution still exhibits some problem $vs.$~existing (loose)
data. Accordingly, however, the alternative of DM being warm has lost
much of its appeal; hydro simulations of SCDEW are still being
planned, to test whether a small $m_w$ risks to over--solve previous
\LCDM problems, below the average mass scale, as well as {\it to
  verify the effects of higher baryon spectra, in respect to DM ones},
as predicted by low--$m_w$ SCDEW cosmologies.

Within this context, models with greater $m_w$ values are an open
option. In order to obtain BH seeds with a mass $\sim 400\, M_\odot$,
a value of $m_H \sim 2\, $keV should be favored. If one prefers not
to invoke a later accretion at the Eddington limit, a value $m_H \sim
1\, $keV might be preferable.

According to PS or similar mass functions, at low mass scale, the
ratio between the expected numbers of systems forming with different
masses $N(>M_1)/N(>M_2) \sim M_2/M_1$. Let is then evaluate this ratio
for $M_1 \simeq 400\, M_\odot$ and $M_2 \simeq 4 \times 10^8 M_\odot$
(a minimal galaxy scale); we find that the number of candidate BH
seed, for each galaxy, is obtainable by multiplying the ratio $\sim
10^6$ by the $c/a > 0.99$ likelihood $\sim 2 \times 10^{-4}$. In this
case, we therefore obtain 200 ``possible'' PBH per galaxy. This kind
of estimate allows us to rise $m_H$ up to $\sim 1\, $keV.


Of course, for such a wider number of expected PBH, the question of
them causing early reionization is to be deepened. In turn, it is
clearly arbitrary to guess that 1:200 ``candidate'' PBH turn into
actual PBH.


\section{Conclusions}
SCDEW models are likely to be the possible physics underlying the
successful \LCDM paradigm, eliminating all conundrums due to $\Lambda$
and even allowing for the same field being both DM and inflaton.  In
this paper we show that their spectra are close to \LCDM also in the
high--$k$ range. 

The main point made here is that, in top of that, SCDEW predicts
primordial BH with a mass--scale up to $\sim 10^7$--$10^8 M_\odot$, a
value reached if present DM particles have a mass $\simeq 100\,
$eV. The recent success of \LCDM hydro simulations, in explaining data
previously met by SCDEW N--body simulations with such $m_w$, allows
for $m_w=m_H$ values exceeding 100 eV. For instance, if $m_w \sim
1$--2$\, $keV, SCDEW predicts PBH in the $10^2$--$10^4\, M_\odot$
range.
In principle, their expected abundance is consistent with the number
of ``seeds'' yielding observed BH's at the center of early and/or late
galactic systems.

\section*{Acknowledgments}

Francesco Haardt is thanked for wide discussions, namely on the
question of early cosmic reionization. We also wish to thank Javier
Rubio for discussions.


\appendix
\section{Evolution of overdensities}

In order to follow the evolution of spherical top--hat overdensities,
we made use of a system of equations obtained by perfecting the
approach followed in \cite{bm2,bm3}.

As done there, let $c$ be the comoving top--hat radius of a coDM
overdensity.  As explained in Section 3, we however try to take into
account other components as well. To this aim, we consider a set of N
concentric shells of wDM ($w$) and baryons ($b$) with comoving radii
$b_n$ ($n=1,...,N$) such that $b_N(\tilde \tau)= c(\tilde \tau)$ at
the initial (conformal) time $\tilde \tau$ (here below, all ``tilded''
quantities will refer to the initial time $\tilde \tau$).

In strict analogy with \cite{mainini2005} and \cite{bm4}, the
evolution equations for the radii read:
\begin{align}
\nonumber
\ddot c~~ &=-\left( \frac{\dot a}{a} -C\dot\Phi \right)\dot c 
-G\frac{\Delta {\cal M}_{co}}{ac^2} \\
~~\ddot b_n &=-\frac{\dot a}{a}\dot b_n 
-G\frac{\Delta {\cal M}_{b_n}}{ab_n^2}
  \label{eq1}
\end{align}
where:
\begin{align}
\nonumber
\Delta {\cal M}_{co}&=\gamma\Delta M_{co}(<c) + \Delta M_{w}(<c) +
\Delta M_{b}(<c) \\
\nonumber
\Delta {\cal M}_{b_n}&=\Delta M_{co}(<b_n) + \Delta M_{w}(<b_n) +
\Delta M_{b}(<b_n)
\end{align}
and 
\begin{equation}
\Delta M_i(<r)=M_i(<r)-\left< M_i(<r) \right> 
=\frac{4\pi}{3} \rho_{cr}\Omega_i\left(\Delta_i-1\right)a^3r^3
  \label{eq2}
\end{equation}
is the mass excess of the $i$ component within a sphere of comoving
radius $r$, $M_i(<r)$ and $\left< M_i(<r)\right>$ being the actual
mass and the average mass respectively.

Let then $x=c/\tilde c$, $y_n=b_n/\tilde c$ and $u=\tau/\tilde\tau$;
the above equations can then be recast in the more convenient form,
suitable to numerical integration:
\begin{equation}
  \nonumber
  x^{\prime\prime} = -h_0 x^{\prime} - \left[\gamma h_{1,co}\left(\tilde \Delta_{co}-x^3 \right)+h_{1,wb}
\left( \tilde \Delta_{wb}-x^3\right) \right] \frac{1}{u^2x^2}
\end{equation}
\begin{equation}
y_n^{\prime\prime} = -\frac{a^{\prime}}{a}y_n^{\prime} -
\left[h_{1,co}\left(\tilde \Delta_{co}-y_n^3 \right)+h_{1,wb}
\left( \tilde \Delta_{wb}-y_n^3\right) \right] \frac{1}{u^2y_n^2} ~;
  \label{eq3}
\end{equation}
here $^\prime$ indicates differentiation with respect to $u$.  The
coefficients
\begin{equation}
\nonumber
h_0=\frac{a^\prime}{a}-C\Phi^\prime \quad, \quad h_{1,i}=\frac{1}{2}\Omega_i h_2
\quad , \quad h_2= \frac{8\pi}{3} G\rho_{cr}a^2\tau^2~,
\end{equation}
while the total density contrast for wDM and baryons:
\begin{equation}
\nonumber
\Delta_{wb}= \frac{\Omega_w\Delta_w + \Omega_b\Delta_b}{\Omega_{wb}} \quad, \quad
\Omega_{wb}= \Omega_w+\Omega_b
\end{equation}
which initial value can be obtained in analogy to (24) for coDM.

It is worth mentioning that the dependence on $\tilde \tau$ of the results 
is carried by the time dependence of the dynamical coefficients 
$h_0$, $h_{1,i}$ and $h_2$.
In the early CI expansion they keep constant and results do not show any 
dependence on $\tilde \tau$; this regime is however abandoned as soon as 
the higgs screening becomes efficient.

\section{Sphericity likelihood}

The scope of this Appendix is determining the expected {\it
  a}--sphericity of fluctuation entering the horizon, as a function of
the factor $F$ by which the fluctuation amplitude exceeds average.

The average {\it a}--sphericity is expected to decrease with $F$.
However, the fluctuations characterized by a given $F$ exhibit an {\it
  a}--sphericity distribution. If we fix suitable {\it a}--sphericity
thresholds $x$, our aim is finding which fraction of fluctuations lay
inside $x$, as a function of $F$.

To do so, we follow a pattern close to \cite{PH}, adding the extra
ingredients needed to reply the above question.  Another important
work, in this field, is \cite{BBKS}.

In the proximity of a maximum, the density fluctuation $\delta ({\bf
  r})$ can be approximated, to the second order, as
\begin{equation}
  \label{nearmax}
  \delta(x_1,x_2,x_3)=\delta_{m}+\frac{1}{2}\left(\delta_{11}^{''}x^{2}+
  \delta_{22}^{''}y^{2}+\delta_{33}^{''}z^{2}\right)~.
\end{equation}
Here $\delta_{m}$ is the density contrast at maximum, while the second
derivatives $\delta_{ii}^{''} = \partial^2 \delta/\partial x_i^2$
($i=1,2,3$), in the directions of the principal axes of the
fluctuation, are evaluated at its maximum (such $\delta_{ii}^{''}$
must be negative to have a confined fluctuation).  The three semi-axes
$a_i$ of this ellipsoidal density distribution read then
\begin{equation}
a^{2}_i=-\frac{2\delta_{m}}{\delta_{ii}^{''}}~.
\end{equation}

If $\delta$ is fully unsmoothed, when extending our sight far from the
maximum, the shape of the fluctuation shall hardly be given by
eq.~(\ref{nearmax}). Being interested in the fluctuation symmetry on
the horizon scale $R_H$, therefore, we need to assume $\delta$ to be
suitably smoothed, so hiding detailed features on scales $\ll R_H$.

Accordingly, our result concerns the basic symmetry of the whole
fluctuation and any information concerning its profile is lost.
Admittedly, they can be important in defining the expected evolution
of fluctuations, after entering the horizon (see, e.g.,
\cite{germus}).

Furthermore, if we suppose to proceed through a gradual increase of
the smoothing ratio, to finally reach a scale $\sim R_H$, we must
expect that the directions of the principal axes can (gradually)
rotate in space. This means that, even when we expect $\delta({\bf r})$
{\it a}--sphericity not to exceed a suitable limit, for the
fluctuation taken as a whole, some {\it a}--sphericity may exist
inside the structure, with compensations among different scales $\ll
R_H$.

As we aim to correlate sphericity to virialization, we have to bear in
mind that, at any {\it a}--sphericity level, inner structures may
frustrate the effects of an apparent overall sphericity. In spite of
that, it is clear that the basic feature to discriminate between
virialization or total collapse is the overall sphericity level.

Let us then outline that, for a gaussian
noise, $\delta_{m}$ and $\delta_{ii}^{''}$ ($i=1,2,3$) follow a
multivariate normal distribution, whose covariance matrix elements are
given by \cite{PH}, and depend on 
\begin{equation}
  \sigma_{n}^{2}(R)=\frac{1}{2\pi^{2}}\int_{0}^{\infty}
  k^{2\left(n+1\right)}P_{\delta}(k)\,W^{2}(k\,R)\,dk
 \label{sigman}
\end{equation}
with $n=1,2,3\, $ (let us outline that these are not vector
components). Here $R$ is the smoothing scale (close to the Hubble
radius $R_{H}$), $P_{\delta}(k)$ is the spectrum of fluctuations, $k$
being the wave number, while $W(kR)$ is the window function used.

For the sake of simplicity, let us then assume that fluctuation
entering the horizon have a spectral index $n_{s}=1$. This forces us
to adopt a gaussian window, to avoid $\sigma_{n}^{2}$ possible
divergences, in spite of dynamical elaluations being based on a
top--hat profile. According to the definition (\ref{sigman}) it shall
be
\begin{equation}
  \label{proptoR}
  \sigma_{n}\propto R^{-n}~,
\end{equation}
so that the density field exhibits an $R$--independent standard
deviation $\sigma_0$.

To generate a column vector {\bf u} whose elements are
$\left(\delta_{m},\delta_{11}^{''},\delta_{22}^{''},\delta_{33}^{''}\right)$,
we use Cholesky decomposition of the covariance matrix $|\mathbf{V|}$
to get $\mathbf{u}=|\mathbf{A|}\cdotp\mathbf{v}$ (the mean values are
equal to zero); here $\mathbf{v}$ is a column vector whose elements
are 4 normal deviates with 0 mean and standard deviation 1;
$|\mathbf{A|}$ is a lower triangular matrix such that the product with
its transposed returns $\mathbf{|V|}$. When one or more values of
$\delta_{ii}^{''}>0$, or $\delta_m < 0$, the vector {\bf u} is
rejected.

Owing to eq.~(\ref{proptoR}), the semiaxes $a_i$ turn out to be
$\propto R$, so that $a_i/R_{H}$ ($i=1,2,3$) do not depend on
redshift. Let us then dub $a$ ($c$) the maximum (minimum) $a_i$
semiaxis, $b$ being the intermediate one, and focus on the ratios
$c/a$ and $b/a$.

We then estimate the fraction of perturbations, for a given value of
$F= (1 + \delta_{m})/\sigma_{0}$, with $c/a$ exceeding 4 possible
thresholds (0.9, 0.95, 0.98, 0.99). To do so, we fix $F$ and generate
a large number ($\sim 10^{5}$) of random replicas for the
ratio $c/a$.

The results are shown in Figire~\ref{marinplot}, in the text. Let us
however outline that, also for small values of $F$, a non negligible
fraction of systems appear to be ``almost'' spherical
(i.e. $c/a>0.99$).

\bsp

\label{lastpage}

\end{document}